\documentclass[10pt,journal,compsoc]{IEEEtran}

\usepackage[nocompress]{cite}
\usepackage[ruled,vlined]{algorithm2e}
\usepackage{times,amssymb}
\usepackage{amsmath}
\usepackage{subfigure}
\usepackage[font=small]{caption}
\usepackage[dvips]{graphicx}
\usepackage{url}
\usepackage{color}

\begin{document}

\title{Content Retrieval At the Edge: A Social-aware and Named Data Cooperative Framework}

\author{Lingjun~Pu,
        Xu~Chen,~\IEEEmembership{Member,~IEEE,}
        Jingdong~Xu,~\IEEEmembership{Member,~IEEE,} and
        Xiaoming~Fu,~\IEEEmembership{Sensor Member,~IEEE.}
}

%\markboth{Journal of \LaTeX\ Class Files}%
%{Shell \MakeLowercase{\textit{et al.}}: Bare Advanced Demo of IEEEtran.cls for IEEE Computer Society Journals}

\IEEEtitleabstractindextext{%
\begin{abstract}
Recent years with the popularity of mobile devices have witnessed an explosive growth of mobile multimedia contents which dominate more than 50\% of mobile data traffic. This significant growth poses a severe challenge for future cellular networks. As a promising approach to overcome the challenge, we advocate \emph{Content Retrieval At the Edge}, a content-centric cooperative service paradigm via device-to-device (D2D) communications to reduce cellular traffic volume in mobile networks. By leveraging the Named Data Networking (NDN) principle, we propose \emph{sNDN}, a social-aware named data framework to achieve efficient cooperative content retrieval. Specifically, sNDN introduces Friendship Circle by grouping a user with her close friends of both high mobility similarity and high content similarity. We construct NDN routing tables conditioned on Friendship Circle encounter frequency to navigate a content request and a content reply packet between Friendship Circles, and leverage social properties in Friendship Circle to search for the final target as inner-Friendship Circle routing. The evaluation results demonstrate that sNDN can save cellular capacity greatly and outperform other content retrieval schemes significantly.
\end{abstract}
}
\maketitle

\IEEEraisesectionheading{\section{Introduction}\label{sec:introduction}}
Mobile devices such as smartphones currently gain enormous popularity and become an indispensable part of our life. For example, recent statistics reflect that daily usage time of smartphones is 4 hours on average, a 35\% increase in time spent from 2014 \cite{Yahoo} and over 90\% of smartphone users have used their smartphones throughout the day since 2012 \cite{Statistics2012}.
In this context, more and more users prefer to exploit their mobile devices to access rich multimedia content on the internet.
Cisco forecasts that mobile multimedia traffic will grow at a compound annual growth rate of 66\% between 2014 and 2019, the highest growth rate of any mobile application category, and that it has accounted for more than 50\% of global mobile data traffic since 2014 \cite{Cisco}. Nevertheless, this significant growth will pose a severe challenge for future cellular networks \cite{chen2013social}. Indeed, a recent report from Small Cell Forum states that network operators can expect their cellular network capacities to increase 29\% on average per year in terms of jointly increasing available spectrum and the number of cell sites as well as improving network performance \cite{FemtoForum}. As a result, there is a significant gap between user increasing demand for multimedia and the supplied capacity of cellular networks (i.e., 66\% vs. 29\%), which begs for further innovations.

Besides increasing the cellular network capacity directly, there are several ideas try to narrow such a demand-supply gap.
The first is to consider data usage based price plans to limit heavy cellular traffic. However, pure usage based plans are likely to backfire by a particular section of user groups such as young smartphone users who have the highest potential for the future revenue growth \cite{SmartphoneUse}. Therefore, it is complicated to design optimal price plans jointly considering cellular traffic, operators revenue and user friendliness.
The second is to consider WiFi (or Femtocell) offloading (see \cite{aijaz2013survey} as a survey) since building more WiFi hot spots is significantly cheaper than network updates for network operators. Although many operators and communities are starting to integrate WiFi to their cellular cores as a heterogeneous network (HetNet) \cite{FemtoForum}, they face a lot of challenging issues such as where and how much WiFi spots should be deployed in terms of service coverage, and how many mobile users should be attached to a WiFi spot in terms of service quality (e.g., ensuring a basis user transmission rate) \cite{Ding2015}.

In parallel with the preceding ideas, an emerging service-oriented topic called mobile edge computing attracts broad attention from both industry and academia.
Mobile edge computing enables a collaborative multitude of end-user and/or near-user edge facilities (e.g., base stations and cloudlets) to cooperatively carry out a substantial amount of services they can each benefit from (see \cite{MEC_2014} as a survey). In this spirit, we advocate a novel cooperative content\footnote{We consider the meaning of ``content" equals to that of ``data", and will use them interchangeably throughout this paper.} retrieval at the edge via device-to-device (D2D) communications. Its rationality is four-fold. First, content (e.g., videos) popularity normally follows a power-law distribution or zipf distribution \cite{cha2009analyzing}, which indicates that hot contents are interested by a lot of users. In addition, many large-volume hot contents such as online course videos are delay-tolerant \cite{ra2010energy}. Third, at network edge there are sufficient users with content stored in their mobile devices to form an ad-hoc ``content cloud" \cite{liu2013exploring, sermpezis2014not}.
At last, D2D communications such as WiFi-direct and LTE-D2D \cite{liu2015device} are promising to replenish traditional cellular communications in terms of user throughput increase, cellular traffic and monetary cost reduction and network coverage extension \cite{asadi2014survey}. As such, cooperative content retrieval empowers mobile content consumers (i.e., users who demand contents) to benefit from the ad-hoc ``content cloud'' so as to meet the growing demand for multimedia, release the pressure on the cellular network and increase the network connectivity\footnote{Mobile users may experience poor cellular connectivity in an area (e.g., at  edge or out of the cellular service range) or are sensitive to the monetary cost of cellular data usage.}.

The main purpose of this paper is to develop an efficient cooperative content retrieval framework for mobile users. A straightforward approach is to flood content request packets and content reply packets \cite{pitkanen2009searching}. However, it would generate high energy overhead due to a tremendously large number of transmitted packets, which is not suitable for mobile devices. As content consumers specify \emph{what} they search for rather than \emph{where} they expect it to be provided, we propose a content-centric and consumer-driven approach by leveraging the Named Data
Networking (NDN) principle \cite{zhang2014named}, a best-known Information-Centric Networking (ICN) solution.
The reason is that cooperative content retrieval is akin to the communications in NDN which are driven by content consumers, through the exchange of two types of packets: Interest (content request) and Data (content reply).
Moreover, mobile devices suffering from limited energy and storage also adapt to NDN single-copy one-to-one Interest and Data matching policy.

A common strategy in existing NDN-based schemes \cite{hoque2013nlsr, garcia2014name} is that users advertise their available contents, build content routing tables from received advertisements, and forward Interest and Data according to those routing tables. In mobile networks, however, this is not feasible since the possible route in content routing tables may be expired due to users' high mobility and intermittent connection \cite{grassi2014vanet}. Since mobile devices are carried by human beings, it is natural to take human social characteristics such as mobility pattern and interest preference into consideration. This inspires us to propose \emph{sNDN}, a social-aware NDN framework to achieve efficient cooperative content retrieval, and we concentrate on answering the following questions:
\begin{enumerate}
  \item What kinds of social characteristics are useful for cooperative content retrieval?
  \item How does NDN integrate with those social characteristics to achieve efficient cooperative content retrieval?
\end{enumerate}

In this paper, we consider a content-centric cooperative service paradigm via D2D communications to reduce cellular traffic volume in mobile networks. In contrast to ``push" services such as host-to-host routings and content disseminations, this ``pull" paradigm where content consumers are unaware of content providers in advance evolves a challenging problem: how to design an efficient content request and reply routing. To the end, we introduce \emph{Named Friendship Circle}
where the centric user and her close friends share high content similarity due to common interest as well as high mobility similarity due to frequent contact.
In this context, we formulate social strength as a combination of these two similarity, then propose a dynamic circle construction and naming policy.
With the ``converged'' and ``stable'' features, Named Friendship Circle can be regarded as both ``data area'' and ``user area'' which respectively divide the content request and reply packet routing into macro and micro steps. For macro steps, we construct NDN routing tables to lead packets to a suitable Named Friendship Circle. For micro steps, we leverage social properties in Named Friendship Circle to steer packets toward the desired targets. We also discuss the practicality of sNDN, and conduct extensive simulations to evaluate its performance. The evaluation results demonstrate that sNDN can save cellular capacity greatly and outperform other content retrieval schemes significantly.

\section{Related Work}\label{sec:related}
The idea of exploiting social concepts to facilitate content forwarding and sharing has been widely studied in opportunistic mobile social networks (see \cite{zhu2013survey, wei2014survey} as a comprehensive survey).

Many existing researches propose social-based routings for efficient host-to-host forwarding. For example, BubbleRap \cite{hui2011bubble} groups mobile users with frequent contacts and ranks them with global social centrality and local social centrality. A packet in BubbleRap is firstly routed towards the users with higher global centrality to reach the destination's community, and further finds the destination with local centrality in the community. However, such host-to-host routings cannot apply for cooperative content retrieval since the destination (i.e., mobile content provider) is unknown in advance. Moveover, traditional community definitions only consider the user contact but neglect the data perspective, which may result in diverse data category in a community.

Alternative directions focus on content-based dissemination.
SocialCast \cite{costa2008socially} calculates a user utility value on an interest based on the user social relationship to the users subscribed to the interest. It publishes contents on an interest to subscribers by forwarding the contents to the users with the higher utilities on the interest.
Opp-Off \cite{han2012mobile} exploits opportunistic communications to facilitate information (i.e., advertisement and software update) dissemination and thus reduces the amount of cellular traffic. It tries to select the minimum number of forwarding users on the condition that the number of ``infected" users (i.e., those users who should receive the information) is over a threshold.
iCast \cite{moghaddam2014interest} is an interest-aware multicast scheme, where contents are proactive sent to the users whose behavioral interest profile matches the contents.
However, these methods mainly focus on disseminating publications to matched
subscribers (i.e., provider-driven), which cannot be applied to cooperative content retrieval (i.e., consumer-driven) directly.

Recently, researchers start to consider content searching in mobile networks. LOAD \cite{shen2014scalable} maps
content metadata to geographical regions and stores the contents in multiple users in the regions. Then, it proposes a novel region-based geographic routing protocol in terms of metadata for content searching. However, it mainly facilitates location-aware content searching (e.g., queries for road congestion and parking information).
The most close to our work are \cite{li2009mops, lu2014information}.
MOPS \cite{li2009mops} provides content-based query/response service by utilizing the social community concept. It groups users with frequent contacts and selects users that connect different groups as gateways.
Then, content requests and replies are relayed through gateways to reach different communities. However, MOPS puts much burden on those gateway users since they have to summarize the interest and  maintain content index of all the users in their connected communities. In our scheme, we utilize the user Friendship Circle concept considering both user content similarity and mobility similarity, and we do not introduce gateway users but let the centric user build her own Friendship Circle to maintain the content name from her close one-hop friends. The most recent research \cite{lu2014information} inspired from NDN schemes proposes an advertisement-based content retrieval method in which the users with higher social centrality exploit Bloom filter to build content synopses when they receive the content advertisements from lower centrality users. A content consumer first gets to know the provider with the desired content by checking content synopses from those encountered users, and then sends a content request packet to the provider via a host-to-host routing. However, the collection and maintenance of content synopses from many users in mobile networks are costly. In addition, this idea may cause long delay since ``knowing" the provider in the first routing step does not mean to ``reach" him in a short while. Our scheme enables the centric user to collect the information from her limited one-hop friends, which is lightweight and practical. What's more, our routing principle is to ``reach" a desire provider without ``knowing" him in advance, which will reduce the content searching time.

There are also several approaches integrate ICN/NDN with mobile networks, while none of them is designed for mobile devices. For example, Grassi \emph{et al.} \cite{grassi2014vanet} design a V-NDN prototype in vehicular networks. They apply vehicle characteristics (e.g., sufficient energy and storage) to update the NDN protocol stack, such as enabling vehicular nodes to cache data packets by one-hop flooding, even though they do not have the matched PIT entries. Since mobile devices suffer from limited energy and storage, the V-NDN scheme is not applicable.

Compared with existing researches, sNDN is a novel framework that integrates user social relationship with NDN principle to facilitate cooperative content retrieval. More specifically, (1) data availability in sNDN is spontaneously maintained within user's Friendship Circle to overcome user mobility issue. (2) We propose a new clustering idea by jointly considering both user content similarity and mobility similarity. (3) We propose a social-aware and named data framework in the context of the Friendship Circle, and evaluate its performance with extensive trace-based simulations. Note that, our framework is not a panacea but a new attempt in the ecosystem of mobile edge computing. We admit that many challenges remain, e.g., optimal forwarder selection and content delay issues. Addressing these concerns is a direction of our future work, and we hope our proposal will provide some potential guidelines to facilitate future content retrieval design at the cellular network edge.

\section{Cooperative Content Retrieval Scenario}\label{sec:scenario}
\begin{figure}[thb]
\centering
\includegraphics[height=1.2in]{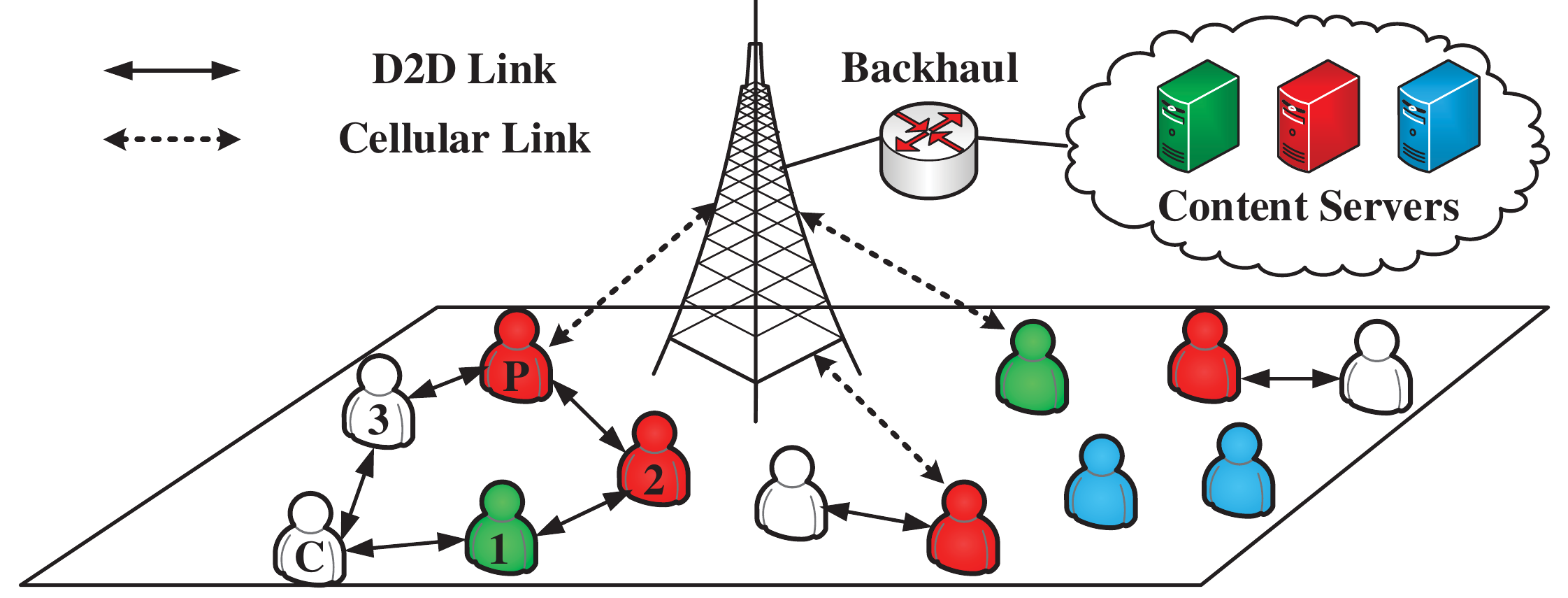}
\caption{An illustration of cooperative content retrieval}
\label{fig:scenario}
\end{figure}
We consider a cooperative content retrieval scenario in a mobile network (see Fig. \ref{fig:scenario}) which includes a set $M$ of mobile
users and a set $D$ of popular data stored in remote content servers.
We assume that content servers can provide a global uniform naming policy such as flat naming for any data $d \!\in\! D$. Note that, designing a perfect naming policy is out the scope of this paper, and we only assume that the name of $d$ is composed of several name components from a global namespace $N$. For example, an online course video can be named by \{``Coursea'', ``Computer'', ``Stanford'', ``Java'', ``Week1"\}. We also assume that an interest can be regraded as a name component $\delta \!\in\! N$, and a data request can be formed by one or more interests. For example, if a user is interested in Java programming and online course from Coursea, she will have a ``Java" and a ``Coursea" interest. Then if she wants to obtain a Java course from Coursea, she will launch a data request with the name \{``Coursea'', ``Java''\}.

As to mobile users, we consider that each user $i \!\in\! M$ has personal interests (e.g., distinct colors in Fig. \ref{fig:scenario}) to request data in content servers. We assume that mobile users can obtain the desired data by directly accessing remote content servers via cellular communications, and some users who are at the edge or out of the service range of a cellular base-station (or sensitive to the monetary cost of cellular data usage) would prefer the cooperative content retrieval to obtain data via D2D communications. The reasons have been given in section \ref{sec:introduction}. In addition, we define a mobile content provider (e.g., user 1) with respect to an interest (e.g., green color)
as a user who has possessed the data covering that interest, and we further define a mobile content consumer
as a user who generates a data request but cannot or is unwilling to obtain the data via cellular communications.

A content consumer can initiate the cooperative content retrieval and seek assistance from nearby mobile users to find the desired data stored in a mobile content provider via D2D communication hops.
Specifically, when a mobile content consumer (e.g., user $C$ launching a data request with respect to red color interest) contacts another user (e.g., user $3$), she will send an Interest packet with the data request and the user who receives the packet becomes a forwarder, and is responsible to help to find the desired data with future D2D contacts. When a mobile content provider (e.g., user $P$) with the matched data receives the Interest packet, he encapsulates the data in a Data packet and launches Data packet routing. With carry-forward mechanism (e.g., user $P\!\rightarrow\!2\!\rightarrow\!1\!\rightarrow\!C$), the Data packet will reach the content consumer in the end.

In this paper, we assume all users are collaborative. However, there are also scenarios that some users are selfish (i.e., refusing to help other users). To motivate the selfish users, we can use some token-based incentive mechanism \cite{chen2010mobicent, ning2013self}, making use of notional credit to pay off users for packet forwarding. In addition, due to lacking of suitable incentive and limited device energy, we assume that cooperative mobile users would not help content consumers to retrieve the desired data from remote content servers via their cellular communications. At last, since users do not have a stringent content retrieval requirement (i.e., delay-tolerant), we assume that if a user activates the cooperative content retrieval, she will set a countdown timer (e.g., the TTL mentioned in section \ref{sec:Eva}). When the timer expires, the user can activate another round of cooperative content retrieval or download the content directly from the servers.

\textbf{Main Challenge:} In contrast to host-to-host routings and contest-based disseminations\footnote{Note that, some kinds of contest-based disseminations (e.g., advertisement or software update disseminations \cite{han2012mobile, moghaddam2014interest}) do not require specific destinations, which are still different from cooperative content retrieval.} where clear destinations (subscribers) are determined in advance, content consumers in cooperative content retrieval are unaware of suitable content providers, and thus cannot decide whether and to whom they should send the Interest packet. Therefore, it comes out the key challenge: \emph{how can we design an efficient Interest packet routing}?

\textbf{Key Idea:} Our design takes inspiration from searching for a book in a library. Intuitively, a library contains many bookshelves each of which has many correlated books (i.e., from the same category). If a user wants to find a specific book (e.g., a Java book), she normally will visit a bookshelf which may contain the desired book (i.e., the category of the bookshelf includes Java), and then finds the desired book %on the layer with the largest number of correlated books
on that bookshelf. Similarly, in order to design an efficient Interest packet routing we should answer the following two questions:
\begin{enumerate}
  \item How to define a stable content area in a mobile network analogous to a bookshelf in a library?
  \item How to visit an area which may contain the desired content, and further find the desired content in that area?
\end{enumerate}
As to the first question, we believe that the Friendship Circle formed by a user and her close friends can be regarded as the ``stable content area''. The rationality is that a user's Friendship Circle is less frequently changed, and the data category is often converged since a user's data taste is similar to and/or influenced by her friends in the circle \cite{yang2012circle}. As to the second question,
we believe Interest name based routing schemes should be considered since Interest name (analogous to book name) helps to decide whether a user's Friendship Circle (analogous to bookshelf) is preferable (i.e., whether the data category of a Friendship Circle covers the Interest name). Therefore, we propose \emph{sNDN}, a social-aware and named data framework\footnote{We also design a Data packet routing for ensuring integrity of services.} for cooperative content retrieval in the following.

\section{Named Friendship Circle}\label{sec:NFC}
In order to exploit Friendship Circle for cooperative content retrieval, in this section we first formulate the social strength between users, then exploit it to construct Friendship Circle, and finally propose a naming policy for Friendship Circle to indicate its data category.

\subsection{Social Strength Formulation}\label{sec:social}
Sociologically speaking, two users achieve a strong social strength if they often share mutual interests and feelings, and repeatedly encounter without making special plans \cite{mcpherson2001birds}. To the end, we respectively leverage user data similarity (i.e., logical strength) and user mobility similarity (i.e., physical strength) to depict these two aspects in mobile networks.

\begin{figure}[thb]
\centering
\includegraphics[height=1.6in]{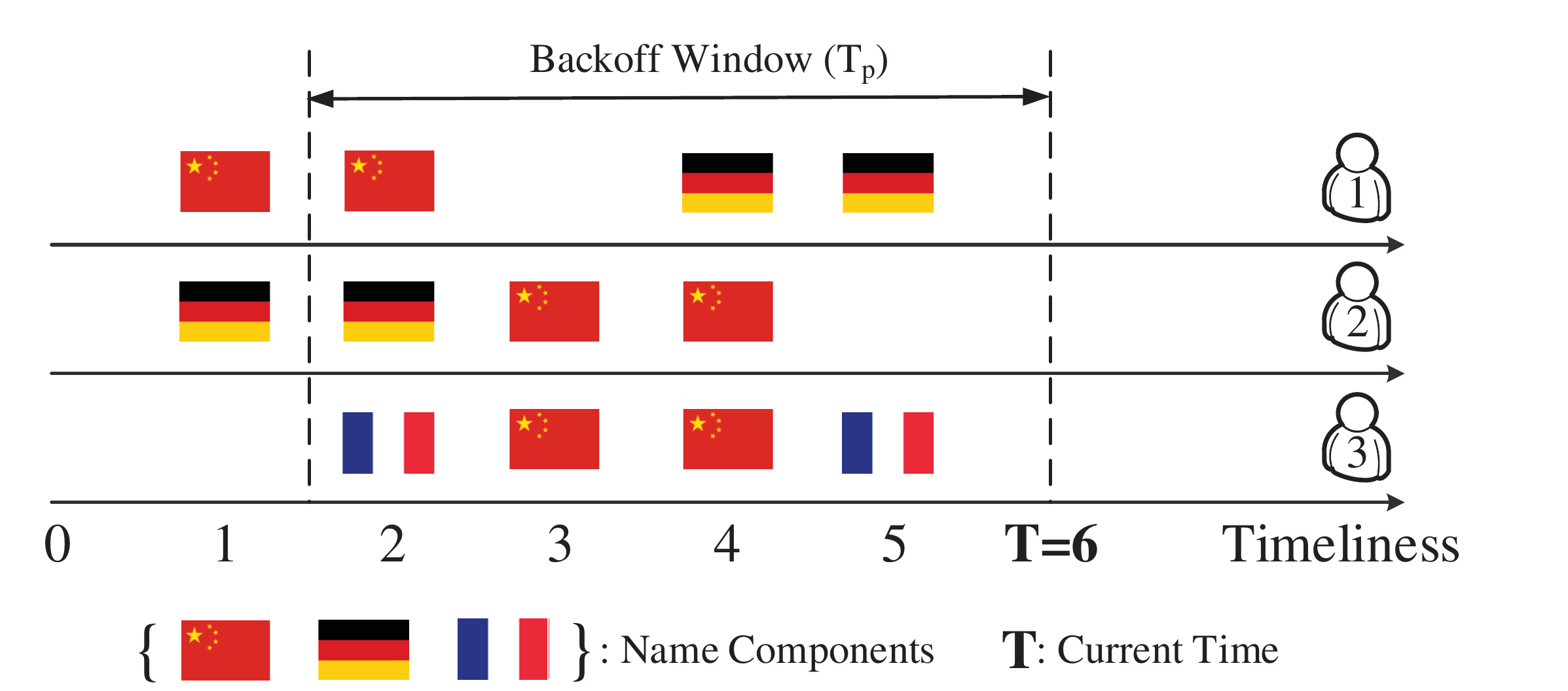}
\caption{A contrived case for logical strength explanation}
\label{fig:data_similarity}
\end{figure}

As mentioned in section \ref{sec:scenario}, each data is identified by multiple name components, and user interests are well captured by name components. Therefore, for the formulation of logical strength, we introduce \emph{component freshness} and \emph{component similarity}. The component freshness reflects the importance of a name component to the time-varying user interest since more fresh name components better depict current user interests and mobile users may delete stale contents to save their device storage in a near future. To reveal the timeliness, we use mathematical convolution to formulate the freshness of a name component $\delta \in N$ in a data obtained by user $i$:
\begin{equation}\label{equ1}
\setlength{\abovedisplayskip}{3pt}
\setlength{\belowdisplayskip}{3pt}
F(i, \delta) = R(\delta, T) \otimes \frac{1}{{{T}}} = \int_0^{{T}} {R(\delta, \tau ) \times \frac{1}{{{T} - \tau }}d\tau },
\end{equation}
where $T$ is the current time and $R(\delta, \tau)$ is a binary value indicating whether there is a data obtained at time $\tau$ that contains the name component $\delta$. An intuitive understanding of equation (\ref{equ1}) is as follows. According to $R(\delta, \tau)$ definition, we consider that $R(\delta, \tau)\!=\!1$ indicates
a user has an interest in terms of $\delta$ with intensity is 1 at a time point $\tau$. As user interest will frequently drift with time, we consider that the
remaining interest intensity at the current time $T$ is inversely proportional to the time interval (i.e., $T-\tau$). Therefore, name component freshness reflecting the importance of a name component to the time-varying user interest can be interpreted as the accumulated remaining interest intensity of that name component from a starting time to the current time.
Consider a contrived case in Fig. \ref{fig:data_similarity} where a national flag represents a unique name component and each user obtains a data whose name includes a national flag at each time (``blank" indicates that a user does not obtain a data at that time). According to the equation (\ref{equ1}), we can have China component freshness of user 1 is
0.45 (i.e., $\frac{1}{6-1}\!+\!\frac{1}{6-2}$), and that of user 2 is 0.83 (i.e., $\frac{1}{6-3}\!+\!\frac{1}{6-4}$). Note that, since the convolution operation ensures that the stale data has little influence on the freshness value, to relieve computation in mobile devices we can only consider data storage history in a recent period of time (i.e., a backoff window with size $T_p$). How the window size $T_p$ affects the overall performance will be discussed in section \ref{sec:Eva}.

The component similarity indicating the logical strength is measured by the common name components integrating component freshness between users $i$, $j$,
which is formulated by the \emph{Jaccard similarity ratio} as follows:
\begin{equation}\label{equ2}
\setlength{\abovedisplayskip}{3pt}
\setlength{\belowdisplayskip}{3pt}
W_{l}(i, j) = \frac{\sum\nolimits_{k = 1}^n{{min\{F(i,\delta_k),F(j,\delta_k)\}}}}{\sum\nolimits_{k = 1}^n{{max\{{F(i,\delta_k)},{F(j,\delta_k)}\}}}},
\end{equation}
where $n$ is the size of name component union of $i$ and $j$ in $T_p$. A higher $W_{l}(i, j)$ suggests that user $i$ and $j$ not only share more similar name components (i.e., user interest), but also the interested time with respect to their common name components is more consistent, and therefore indicates that those two users have stronger logical strength. Note that, the traditional Jaccard similarity ratio which is defined as the size of the intersection (i.e., the number of data with common name components) divided by the size of the union of the sample sets is incomplete to model the local strength. For example, if we adopt the traditional method for the users in Fig. \ref{fig:data_similarity}, then $W_l(1,3)\!=\!W_l(2,3)\!=\!\frac{2}{6}\!=\!0.33$. Intuitively, we can see that the local strength between user 2 and 3 should be larger than that between user 1 and user 3 since they have more consistent interest in terms of China component. When it comes to our method, we can have $W_l(1,3)\!=\!\frac{0.45}{0.83\!+\!1.25\!+\!1.5}\!=\!0.14$ and $W_l(2,3)\!=\!\frac{0.83}{0.83\!+\!1.25\!+\!0.45}\!=\!0.33$, which is in accordance with the intuition. In addition, we should emphasize that our method also admits the users who share many common name components achieves high local strength. For example, $W_l(1,2)\!=\!\frac{0.45\!+\!0.45}{0.83\!+\!1.5}\!=\!0.39$ which is higher than the preceding ones.

Note that, the logical strength model does not require extra storage space. This is because that when smartphone users obtain contents such as downloaded documents or cached videos in applications, the operating system will create corresponding file descriptions and provide API to access them. For example, Android provides MediaStore class to manage all multimedia contents. In addition, we specify that the content obtained time in equation (\ref{equ1}) will be the last time when users use contents (initially, it equals to the content creation time). Therefore, we can easily get the content name (when NDN naming policies are used in the future) as well as the last modified time through API functions.

\begin{figure}[thb]
\centering
\includegraphics[height=1.0in]{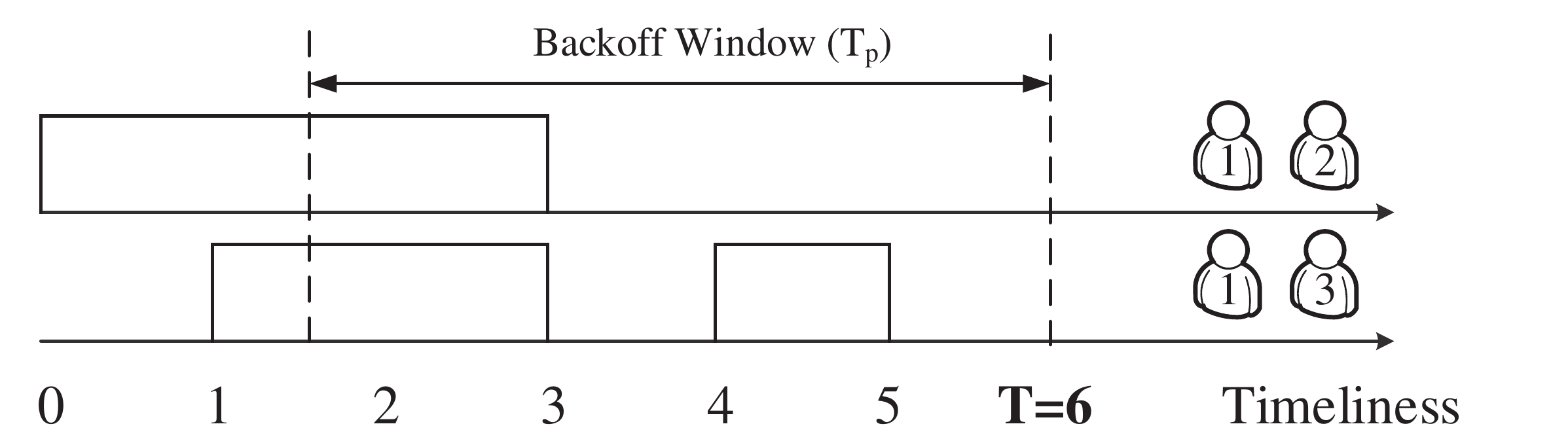}
\caption{A contrived case for physical strength explanation}
\label{fig:contact_similarity}
\end{figure}

The physical strength measures user mobility similarity. It can be derived from user historical contact information such as contact frequency and contact duration time. The former, representing how often two users contact in a period of time, is good at assessing frequent contacts of mobile users with short duration. The latter does well in evaluating the occasional contacts with long duration. Nevertheless, there is no conclusion to say which one is better \cite{zhu2013survey, wei2014survey}. For example,  Consider a contrived case in Fig. \ref{fig:contact_similarity} where rectangle areas represent the user contact duration time and the areas between two rectangles represent the user contact intervals. When we adopt contact frequency to calculate user physical strength, we have the physical strength between user 1 and 2 is less than that between user 1 and 3 (i.e., 1$<$2) while the physical strength between user 1 and 2 equals to that between user 1 and 3 (i.e., 3$=$3) when contact duration time is used. Since these two metrics are orthogonal, in this paper we try to combine them and measure physical strength as a novel \emph{average ratio} that user $i$ will contact user $j$, i.e.,
\begin{equation} \label{equ3}
\setlength{\abovedisplayskip}{3pt}
\setlength{\belowdisplayskip}{3pt}
W_p(i, j)=\frac{{\int_{{0}}^{{T}} {f_{ij}(t)dt} }}{{{T}}},
\end{equation}
where $T$ is the current time and $f_{ij}(t)$ describes the ratio that user $i$ contacts $j$ at time $t$. We set $f_{ij}(t)$ = 1 if two users keep in contact, and $f_{ij}(t) = \frac{1}{t_{next}-t+1}$ otherwise ($t_{next}$ is the next contact time in the historical contact storage). Consider the case in Fig. \ref{fig:contact_similarity}, as to the pair of user 1 and 3 there are 2 contacts $\{(1,3),(4,5)\}$ and 3 contact intervals $\{(0,1),(3,4),(5,6)\}$ before the current time $T\!=\!6$. Then, we can calculate $W_{p}(1, 3)$ as follows:
\begin{align}\label{equ4}
\setlength{\abovedisplayskip}{3pt}
\setlength{\belowdisplayskip}{3pt}
W_p(1, 3) &= \{\int_{0}^{1}{\frac{1}{1-t+1}}dt+\int_{1}^{3}{1}dt+\int_{3}^{4}{\frac{1}{4-t+1}}dt \notag \\
    &+\int_{4}^{5}{1}dt+\int_{5}^{6}{\frac{1}{6-t+1}}dt\}/6 \notag \\
    &=\{ln(1-0+1)+(3-1)+ln(4-3+1) \notag\\
    &+(5-4)+ln(6-5+1)\}/6 \notag\\
    &=\{\Big[(3-1)+(5-4)\Big] + \Big[ ln(1-0+1) \notag \\
    &+ln(4-3+1)+ln(6-5+1)\Big]\}/6 \notag \\
    &=  \{\sum_{n=1}^{2}{t_{{intra}_n}}+\sum_{m=1}^{3}ln(t_{{inter}_m}\!+\!1)\}/T,
\end{align}
where $t_{{intra}_n}$ represents the contact duration of the $n$-th contact, and $t_{{inter}_m}$ is the $m$-th contact interval time. We can see that the equation (\ref{equ4}) reflects the effect of the preceding two metrics and admits users with higher contact duration and frequency to achieve higher physical strength (e.g., $W_p(1, 2)<W_p(1, 3)$). In addition, according to the deduction of equation (\ref{equ4}), users only recording the contact start and end time with other users can facilitate user physical strength calculation, which will not introduce much storage and computation overhead. Note that,
in daily life users can form different Friendship Circles at different time. For example, a student may establish a Networking study circle at working time and a Football circle at leisure time. To reflect the timeliness of Friendship Circle, we require to consider contact storage history in a recent period of time (e.g., we can choose the same time window $T_p$ as that in the calculation of logical strength).

Since users with high physical strength may not share many contents with common categories and vice versa, we define the combination of logical and physical strength as social strength for cooperative content retrieval. That is,
\begin{equation}
{W_{s}(i, j)} \,  =  \, \alpha {W_{l}(i, j)} \,+\, (1 - \alpha ){W_{p}(i, j)},\quad \alpha \in {\rm{[0,1]}},
\label{equ:ws}
\end{equation}
where $\alpha$ determines the weight of the two ratios, $W_{l}$ and $W_{p}$. The impact of $\alpha$ value on the performance is evaluated in section \ref{sec:Eva}.

\begin{figure*}[thb]
\centering
\includegraphics[height=2.5in]{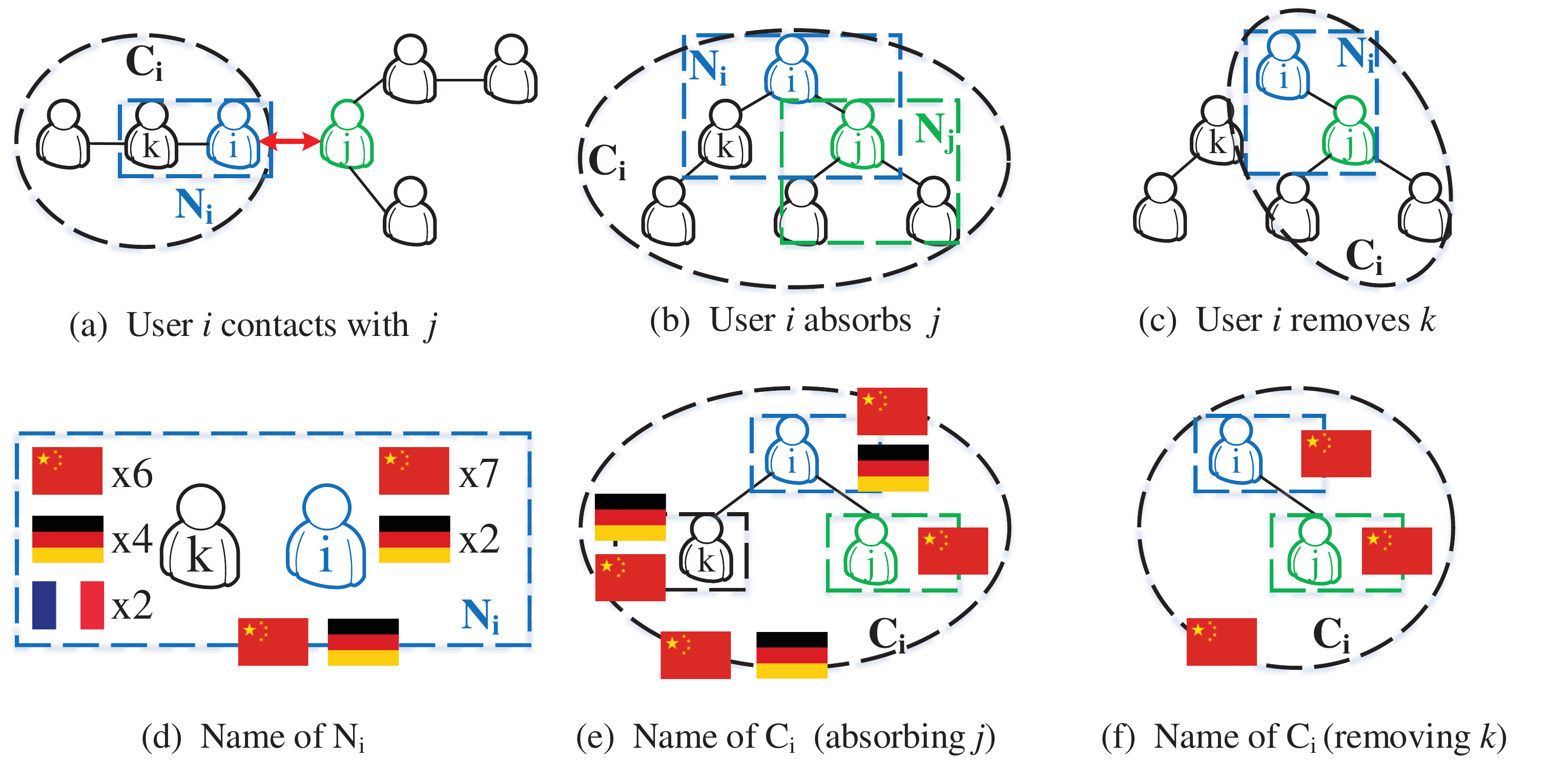}
\caption{An illustration of Named Friendship Circle construction}
\label{fig:FC}
\end{figure*}

\subsection{Friendship Circle Construction}
We next exploit the social strength to construct Friendship Circle. Here, we restrict a user's Friendship Circle to a two-hop structure. This is because, from social aspect, the social tie between a user and her friend's close friends (i.e., one-hop friends) would also be strong \cite{konstas2009social} and from physical aspect, contacting a user would incur long delay if the maximum D2D communication hops are larger than two \cite{li2014can}. To facilitate Friendship Circle construction, each user $i$ maintains two data structures: (1) a neighbour set $N_i$ includes a set of tuples, and each tuple records a one-hop friend $j$ of user $i$ and the associated social attributes $S_j$. (2) A treelike Friendship Circle $C_i$ saving the neighbour set (i.e., $N_j$) of each friend $j$. As illustrated in Fig. \ref{fig:FC}, the procedure with which Friendship Circle is built up is summarised as follows:

\noindent \textbf{Step 1}: Initially, each user {$i \!\in\! M$} sets $N_i$ and $C_i$ to itself.

\noindent \textbf{Step 2}: When $i$ encounters another user $j$, they will exchange limited information to calculate $W_s(i, j)$ (Fig. \ref{fig:FC}a).

\noindent \textbf{Step 3}: If the $W_s$ is higher than a predefined threshold, $i$ will add its counterpart $j$ to $N_i$ and add a new branch $j$ associated with $N_j$ into $C_i$ (Fig. \ref{fig:FC}b).

In step 2, the ``limited information" refers to all name component freshness in the backoff time window.
The reasons are as follows. Firstly, each user in mobile networks requires to execute neighbor discovery service to recognize encountered users, and hence they will keep their own contact history with other users. This has been validated by many realistic user contact traces such as \emph{infocom06} and \emph{MIT Reality}. In other words, when two user contact with each other they do not require to exchange contact history for physical strength calculation. Secondly, according to the equation (\ref{equ2}), users only require to exchange name component freshness rather than detailed content information for logical strength calculation.
In the context of backoff time window, we believe the exchange information is limited and lightweight. In addition, we assume that the users participating the cooperative content retrieving service will truthful share their name component freshness with each other (i.e., their interest preference), which is similar to many preference-aware content dissemination researches in mobile networks\cite{costa2008socially, gao2011user, lin2012preference}.

\begin{figure}[thb]
\subfigure[\small \emph{Infocom06} trace]{
\includegraphics[height=1.2in]{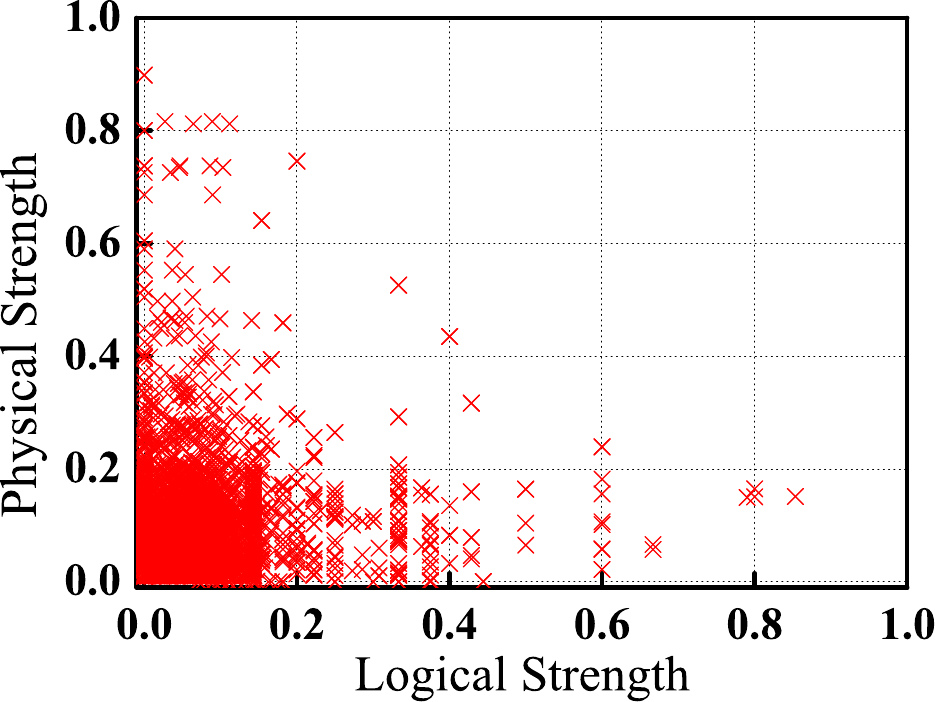}}
\subfigure[\small \emph{MIT Reality} trace]{
\includegraphics[height=1.2in]{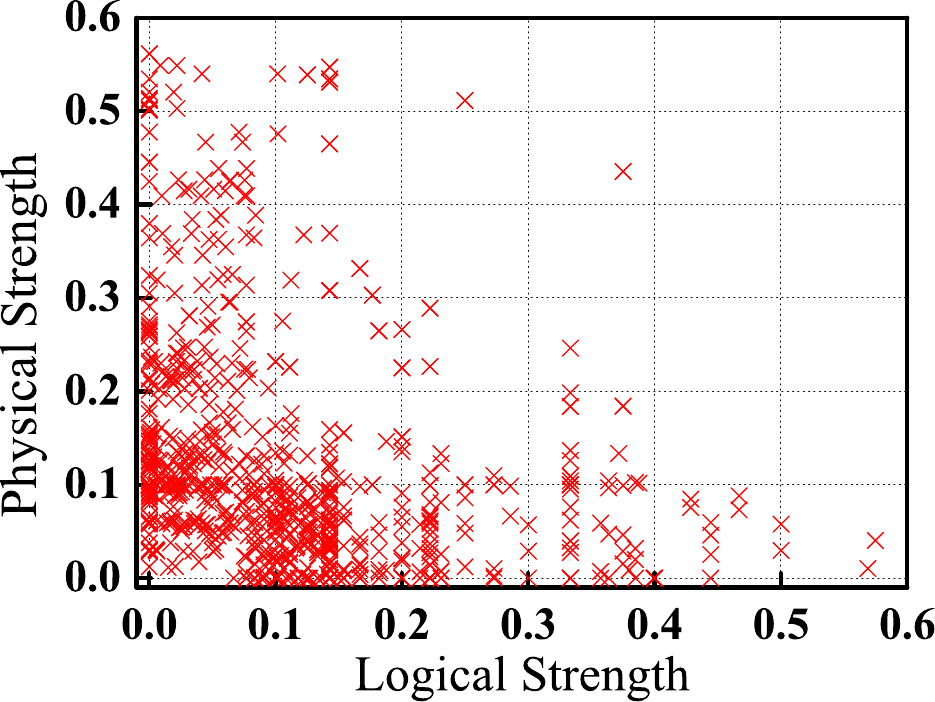}}
\caption{Statistics of the pair of physical and logical strength of the 3th day in the simulation when window size (i.e., $T_p$) is set to 3 hours.}
\label{fig:threshold}
\end{figure}

Since close friends may contact with each other many times in a short period, to relieve the information exchange and computing in mobile devices, it is no need to recalculate $W_s$ if they are in the counterpart's neighbour set already. In addition, a user's Friendship Circle may change over time due to the time-varying user interest and mobility. To capture this practical feature, a fresh timer is added in $S_j$ for each friend $j$ in neighbour set. Since the social strength is calculated in the period of time $T_p$, when user $i$ adds $j$ in her neighbour set, she will set a fresh timer for $j$ starting from $T_p$ and counting down every time unit. If a fresh timer (e.g., for user $k$) becomes 0, $i$ will remove that branch from $C_i$ and the user from $N_i$ (Fig. \ref{fig:FC}c).
We should emphasize that, the predefined threshold in step 2 should be determined by the users together with the other players (e.g., network operators and three-party service providers) in cooperative content retrieval in terms of the number of mobile users and contents in the network. For example, in a sparse mobile network, a larger threshold which generally restricts the users in a Friendship Circle with higher social strength may pose difficulty in building a Friendship Circle due to the low user physical strength. In this paper, we set a mild threshold in the simulation in terms of network type and user amount as well as content amount. For example, we record the pair of physical and logical strength when users contact with each other, and plot them in Fig. \ref{fig:threshold}.  In the case, if users and operators would like to restrict the Friendship Circle to a small one (e.g., in a dense network like \emph{infocom06}), they can set the threshold to 0.18 (roughly 50\% contact points falls into the area 0.18$\times$0.18).
We consider that a suitable threshold can be estimated in an offline manner, and we will also consider a self-adapting setting scheme as a future work.

\subsection{Friendship Circle Naming Policy}\label{policy}
Before we discuss the naming policy for Friendship Circle, let us recall the bookshelf case in section \ref{sec:scenario}. Generally, a bookshelf has many correlated books (i.e., from the same category). In other words, in order to get the category of a bookshelf, we first require to obtain each layer category that possesses a lot of books on that layer) on the bookshelf, and then consider the union of all layer category as the category of that bookshelf.
Analogously, if we view a user Friendship Circle as a bookshelf, we can regard a neighbour set in it as a book layer. Therefore, each user $i$ should first derive the name of her neighbour set $N_i$ (i.e., extracting the popular name components with much associated data) and then formulates the name of Friendship Circle $C_i$ as the union of neighbour set name of all the users in $N_i$.

Specifically, each user $i$ maintains a name component map to indicate her own data where the key is a name component and the value is the amount of data associated with that name component (Fig. \ref{fig:FC}d). When a new friend $k$ joins in her neighbour set $N_i$, user $i$ appends a field in $S_k$ to record user $k$'s name component map (Fig. \ref{fig:FC}d). With the new field, user $i$ can aggregate name component maps from all users in $N_i$ (including herself) as \emph{neighbour component map} and extract the \emph{key name components} from it as the name of her neighbour set. The key name components in the neighbour set are determined by the following two metrics: $f_u(\delta)$ is fraction of users who have at least one piece of data associated with a name component $\delta\!\in\!N$, and $f_d(\delta)$ is the fraction of the number of data with the name component to the total data amount in the neighbour component map. In order to accelerate Interest packet routings we define a name component $\delta$ whose $f_u$ multiples $f_d$ is larger than a threshold $\kappa$=25\% as a key name component. Note that, the value 25\% will guarantee that either $f_u(\delta)$ or $f_d(\delta)$ is greater than 50\%, which indicates that the data associated with the component $\delta$ can be easily found in the neighbour set. Following this principle, we can see that China component in Fig. \ref{fig:FC}d is a key name component since $f_u(\text{china})\times f_d(\text{china})\!=\!1\times\frac{13}{21}\!=\!0.62>0.25$. The impact of threshold $\kappa$ on the performance is also evaluated in section \ref{sec:Eva}.

To facilitate the name of Friendship Circle, a user $i$ appends another field in $S_j$ to record the neighbour set name of user $j$ (i.e., key name components). With this new field, user $i$ can obtain her Friendship Circle name as the union of the neighbour set name of all the users in $N_i$ (Fig. \ref{fig:FC}e). Note that, if the user $i$ removes a friend (e.g., user $k$) in her neighbour set, she will recalculate both the name of neighbour set and that of circle according to the residual friends information (i.e., $S_j$) in her neighbour set (Fig. \ref{fig:FC}f).
In practice, maintaining the neighbour component map and the neighbour set name will not consume much storage. This is because users usually have limited content interests \cite{mcpherson2001birds}, and each user and her close friends in the neighbour set have many similar contents with respect to their common interests.
%We should emphasis that, Friendship Circle has two advantages compared with existing community structure \cite{hui2011bubble} for cooperative content retrieval. The first is that Friendship Circle integrates content similarity into its construction, which enable the data category
%in it to be converged. The second is that Friendship Circle consider user relationship in a timeliness manner, which is adaptive to user time-varying interest and mobility while existing community structures consider user relationship in a time-accumulated manner. We will further discuss Friendship Circle with different community structures in section \ref{sec:Eva}.

\section{Social-aware and Named data Framework}\label{sec:sNDN}
\begin{figure}[thb]
\centering
\includegraphics[height=1.2in]{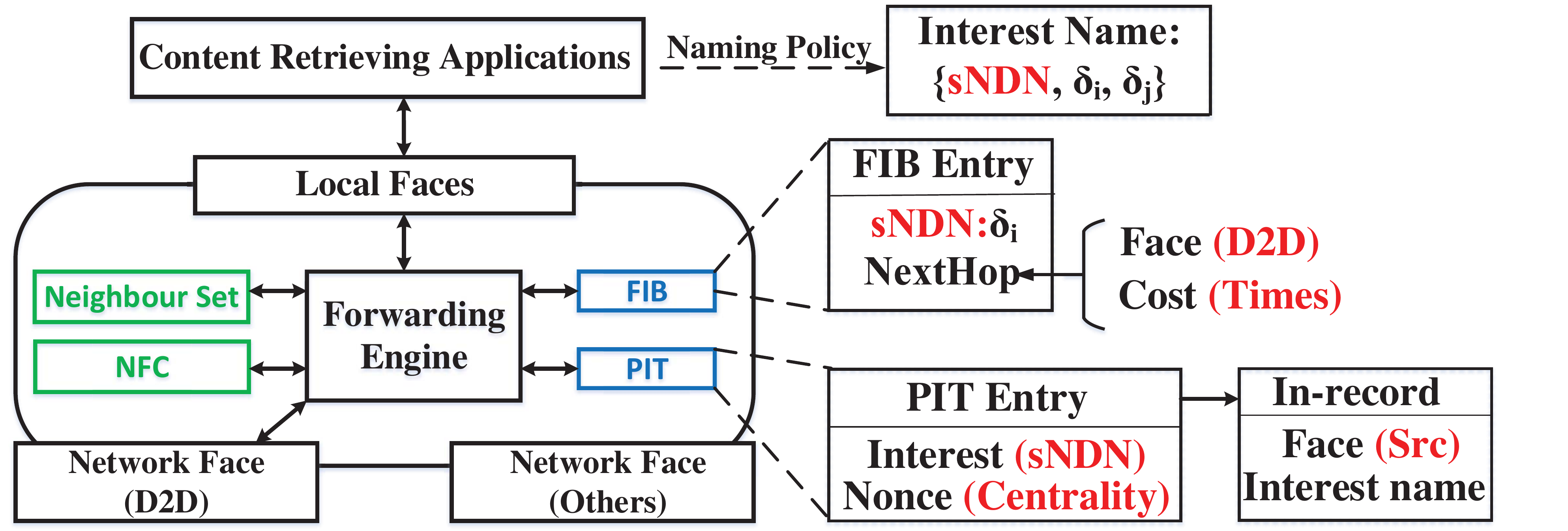}
\caption{An illustration of social-aware and named data framework}
\label{fig:framework}
\end{figure}

In the context of Named Friendship Circle (abbr. NFC), we develop \emph{sNDN}, a social-aware and named data framework for cooperative content retrieval, which is built on the NDN Forwarding Daemon \cite{NFD}, a comprehensive NDN framework. The structure of sNDN is illustrated in Fig \ref{fig:framework} where we attach Named Friendship Circle to Forwarding Engineer, and reuse those fields which are not suitable for packet routings in mobile networks. The main operating procedure is as follows. Applications create Interest packets with the name of requested contents, and put them to the Forwarding Engine through Local Faces. The Forwarding Engine leverages the Forwarding Information Base (FIB) and social data structure (i.e., neighbour component map) to forward Interest packets through D2D Network Faces (e.g., Wifi-direct). Once Interest packets reach mobile content providers with desired contents, the providers will return Data packets with the contents. Then, the Forwarding Engine leverages the Pending Interest Table (PIT) and social data structure (i.e., neighbour set) to forward Data packets back to the requesting applications. Next, we will discuss Interest and Data packet in detail.

\subsection{Interest Packet Routing}
In order to use cooperative content retrieval service, the name of an Interest packet requires to incorporate the prefix ``sNDN". For example, if a user wants to adopt cooperative content retrieval to obtain a Java course from Coursea, she will generate an Interest packet with the name \{``sNDN", ``Coursea'', ``Java''\}. When such an Interest packet is forwarded to the Forwarding Engine through Local Faces, the Interest packet routing of sNDN is invoked.
As mentioned in section \ref{sec:NFC} that Named Friendship Circle indicates ``stable content area'', similar to finding a book in a library, the Interest packet routing can be divided into two steps: an Interest packet is firstly transmitted to a NFC whose name covers the Interest name, and then searches for
the desired data in it.

To facilitate the first step, the Forwarding Engine builds FIB entries for each user depending on the NFC name she encounters. Specifically, the FIB entry in NDN Forwarding Daemon has two fields: Name component and Nexthop. The latter indicates where an Interest packet will be forwarded if the packet name includes the former. However, this field is useless in mobile networks due to user mobility. Therefore, we try to reuse it to design a sNDN-oriented FIB entry. Firstly, we specify that the Name component field requires to incorporate the prefix ``sNDN" to distinguish with normal FIB entries. Then, the face field in Nexthop which used to indicate the specific router interface stores the D2D communication interface in sNDN (e.g., Bluetooth, Wifi-direct and LTE-D2D). At last, we reuse the cost field in Nexthop to save the number of users whose NFC name include the Name component (i.e., count record). As to a mobile user, FIB entries initially are built with her own NFC name and each count record is set to 1. When she contacts with another user (i.e., they are in the mutual D2D communication range), besides executing the Named Friendship Circle construction they exchange their NFC name to update FIB entries (i.e., adds new entries or updates the corresponding entries' count records). Intuitively, as to a name component, a user with higher count record in her FIB has more chance to meet a user whose NFC name covers that name component.
Therefore, we propose a \emph{data directionality} metric for Forwarding Engineer to navigate Interest packet:
$
U_I=\text{min}\{r| \delta \!\in\! I, (\delta, r) \!\in\! \text{FIB} \},
$
where $I$ is the set of name components in Interest name, and $r$ is the count record in terms of the name component $\delta$ in FIB. Note that, if a name component in Interest name does not exist in the FIB, $U_I$ will be 0. Towards the ascending of data directionality, an Interest packet will traverse to a suitable user whose NFC name covers the Interest name.

When an Interest packet enters a suitable user's NFC, Forwarding Engineer will exploit the \emph{neighbour component map} (abbr. NCM) used for Friendship Circle naming to further shrink the search scope to neighbour set (analogous to searching for a book from bookshelf to book layer). Intuitively, as to a name component, a user whose neighbour set has more relevant data will find the desired data more quickly. Therefore, we propose a \emph{data location} metric:
$
U'_I=min\{c| \delta \!\in\! I, (\delta, c) \!\in\! \text{NCM} \},
$
where $I$ is the set of name components in Interest name, $c$ is the data amount
in terms of the name component $\delta$ in neighbour component map. Towards the ascending of data location, the Interest packet
will finally hit a mobile content provider with the desired data.

\subsection{Data Packet Routing}
Once an Interest packet is delivered to a mobile content provider with the desired data, the provider will encapsulate the matching data in a Data packet and send it back. Note that, Named Friendship Circle can also be regarded as ``stable user area'' since the close friends are of frequent contact. Therefore, the Data packet routing firstly locates a user whose NFC includes the content consumer, and further finds her in that circle. Since the reverse path of Interest packet routing cannot be used as Data packet routing in mobile networks, we propose a host-to-host Data packet routing for cooperative content retrieval.
Note that, since host-to-host routings has been widely studied in mobile networks, the purpose of this subsection is to integrate existing idea with Named Friendship Circle and NDN framework to ensure integrity of cooperative content retrieval.

In order to implement a host-to-host Data packet routing, a key issue is that NDN Interest and Data packets do not provide a field to record the identity of content consumer. To solve this problem, we leverage the face field\footnote{In wired stationary networks, the face field records the previous hop node from which Interest packets are sent, which facilitates the reverse path building. However, the reverse path is ineffective in mobile networks, and thus we use the face field to record the consumer identity.} of in-record in PIT entry to keep the identity of data consumer as shown in Fig. \ref{fig:framework}.
Traditionally, the face field will be updated hop by hop to build the forward path, while this field in sNDN will not be updated. In other words, this field will keep the content consumer if the Interest name field of in-record includes ``sNDN". When a content provider receives an Interest packet, he can check the face field of in-record to obtain consumer identity, and then launches a host-to-host Data packet routing smoothly.
As to the routing metric, We adopt the widely used social centrality concept \cite{hui2011bubble} to achieve the first step routing. To the end, we create a special PIT entry with the dummy ``sNDN'' name component to distinguish with PIT entries of other NDN protocols, and exploit the Nonce field to record the social centrality. Towards the ascending of social centrality, a Data packet will traverse to users whose NFC includes the content consumer.

When a Data packet enters a user's NFC, we exploit the \emph{neighbour set} to further shrink the search scope. This is because neighbour set is a one-hop structure and the users in it have a high contact chance. Therefore, the Data packet will be transferred to a user whose neighbour set includes the content consumer, and that user will hold the packet until hitting the consumer. Note that, since data cache itself is a big topic in mobile networks (e.g., related to device storage, user interest preference and data popularity \cite{costa2008socially}), we allows the forwarders not to cache the data in Data packets, and we will consider an efficient caching policy to facilitate sNDN in a future work.

\subsection{Discussion}
In this part, we will discuss the practicality of sNDN by answering the following questions.

\emph{1) What is the energy consumption of ongoing device discovery}:
As mentioned in \cite{trifunovic2013slicing}, for automatic discovery in opportunistic networks, a device can continuously be discoverable
without wasting much energy. However, the process of actively scanning for peers naturally consumes much more energy, which makes discovery scheduling become critical. As to the scheduling issue, we can consider Zheng's and Hou's asynchronous activation algorithm
\cite{zheng2003asynchronous} to check for contact opportunities periodically. The activation pattern within each period is deliberately designed
so that two unsynchronized devices have overlapping activation times. The authors mention that it is possible to
achieve a 9/73 duty cycle, and thus the hourly energy cost for device discovery is $E = P_{idle}\times3600\times\frac{9}{73}$, where $P_{idle}$ is the idle power consumption for staying awake in one D2D mode (e.g., 155.8mW and 340.89mW for Bluetooth and Wifi-direct, respectively \cite{trifunovic2013slicing}). If we consider the mainstream device battery (2000mAh and 3.6V), the $E$ in terms of bluetooth only takes up 0.7\% (i.e., $\frac{155.8\times9}{2000\times3.6\times73}$) of total energy.

\emph{2) What is the energy consumption of D2D communication}:
Communication happens upon each contact, causing a sequence of data transmission with respect to Named Friendship Circle construction as well as Interest and Data packet forwarding.
We should emphasize that, since it is difficult to calculate the exact energy overhead, the  in this paragraph we aim to convey that sNDN will not cause severe energy issue in a high-level view, and leave the fine-grained evaluation in the future work. To the end, we first conduct a simple test to obtain an energy benchmark to facilitate the following statements. Specifically, in the test two smartphones which have closed all the background services, cellular link and screen are connected with each other through Bluetooth or WiFi-direct. One smartphone with full battery will repeatedly transfer a 500Mb file to the other one until its battery is depleted. After the test, we can find that the iterated file transfer is 76 by using Bluetooth and 50 by using WiFi-direct. In other words, transferring 500Mb file between two smartphones using Bluetooth and Wifi-direct will consume 1/76=1.3\% and 1/50=0.2\% of total energy, which is viewed as the benchmark.

As to Named Friendship Circle construction, users require to exchange their name component freshness for circle construction (and their name component map and neighbor set name for name construction if their social strength is high enough) while only the fresh records (i.e., from $T\!-\!T_p$ to $T$) are required to exchange as mentioned in section \ref{sec:social}, and the size of neighbour component map and neighbor set name will not be
too large as mentioned in section \ref{policy}. Therefore, if we assume that the total data exchange for Named Friendship Circle construction is roughly 10M in one communication, the required communication energy in terms of bluetooth only accounts for 0.026\% (i.e., $\frac {10\times1.3\%}{500}$) of total energy.
As to Interest packet forwarding, the energy consumption is limited since the size of Interest packet is quite small \cite{NFD}. As to Data packet forwarding, the requested data in practice will be divided into multiple Data packets which are sent back with different forwarder paths. In addition, some user contact trace analysis \cite{karagiannis2010power, liu2013exploring} mention that the average inter contact time between mobile users spans from minutes to hours, and contact duration lasts for minutes. If we assume the inter contact time is five minutes and each contact duration will transfer five data Packets with size 50MB, then we consider that the hourly energy consumption is still acceptable for a Data packet transmission (i.e., $\frac{12\times5\times50}{500}\times1.3\%=7.8\%$ by using bluetooth). Besides, we admit that the above discussion is a little idealistic (e.g., do not consider the communication distance). We are creating a prototype system in Android smartphones based on an open-source Android DTN application \emph{Bytewalla}\footnote{https://github.com/h0gar/Bytewalla.}, and will evaluate the computation and communication overhead of sNDN in various network scenarios (e.g., office, building, campus) in the future.

%\textcolor[rgb]{1.00,0.00,0.00}{\emph{3) Does sNDN will put much burdens on smartphone storage}: Generally speaking, smartphone users will store their interested contents (e.g., video and music) as well as delete outdated contents depending on their time-vary interest preference. Our proposed framework will not violate user willingness (e.g., enforce them to cache their uninterested or stale content) but to leverage their current cached contents for services. In addition, we consider the user contact history and Named Friendship Circle structure will not require much storage space since the number of close friends or contacted users is limited in daily life. At last, as to the packet forwarding, we consider that the users with high social centrality require to forward more packets for others in current routing scheme, which is not fair for them. Therefore, we will take user incentive and caching issue into routing design in a near future work.}

\emph{3) As we consider a mobile edge computing service, can we integrate network operators into sNDN}: The answer is positive, and we can imagine an \emph{operator-assisted sNDN} in which network operators enable their base stations to facilitate user Named Friendship Circle construction. The rationality is
as follows. To begin with, the base station has the whole network information (e.g., it can execute the device discovery process for each user to detect the set of its nearby users), which benefits user physical strength calculation. In addition, compared with the strangers passing by, users would like trust network operators, and hence network operators can maintain and periodically update user information (e.g., name component map), which benefits user logical strength calculation. What's more, with the centralized control of network operators, the threshold parameters for circle construction can be dynamically updated in terms of time-varying network condition. We will consider the operator-assisted sNDN in a near future work.
Note that, we should emphasize that the network assisted system has been adopted in the emerging 5G networks (e.g., the base station can manage both cellular and D2D connections of mobile users \cite{liu2016outage}) and software-defined mobile networks (e.g., the base station can be viewed as the SDN controller running a collection of application modules, such as radio resource management, mobility management \cite{pentikousis2013mobileflow}).

\section{Performance Evaluation}\label{sec:Eva}
\emph{1) Simulation Scenario}:
Since there is no available dataset involving both user D2D contact records and user content retrieval (or interest preference) records, we attempt to simulate the scenario as discussed in section \ref{sec:scenario}. In order to reflect user contact pattern in diverse mobile networks, we consider two real traces\footnote{Infocom06 and mit dataset. http://crawdad.cs.dartmouth.edu.}: \emph{Infocom06}, a dense conference network and \emph{MIT Reality},
a sparse campus network. These two traces record contacts among users with periodically neighbor discovery. The characters of two traces are shown in Table 1. Note that, we only choose four consecutive days with the highest number of contact records in \emph{MIT Reality} in the following simulation since its daily contact frequency is too small to form a Friendship Circle in most of the time.
\begin{table} [thb]
\centering
%[t]
{\small
\begin{tabular}{|{c}|{c}{c}|}
%\begin{center}
  \hline
        \emph{Traces}  &   \emph{Infocom06} & \emph{MIT Reality} \\ \hline
        %Environment  & Conference & Campus \\
        Network Type & Bluetooth & Bluetooth \\%\hline
        No. Devices &  78 & 97  \\%\hline
        No. Contacts & 191,336 & 54,667  \\%\hline
       % No. links    & 395 &96  & 1,991\\
        Duration (days) &4 &246  \\ %\hline
        %Granularity (secs) &120 &300  \\ %\hline
        Daily Contact Frequency &7.862 & 0.024  \\
%        Contact rate & 0.024 & 0.10& 0.013\\
             \hline
      %  Size ($km^2$) &N/A&358&2367&100\\\hline
%\end{center}
\end{tabular}
}
\caption{Statistics of the user contact traces}
\end{table}

\begin{figure}[thb]
\subfigure[\small Artist distribution in terms of interested user amount]{
\includegraphics[height=1.2in]{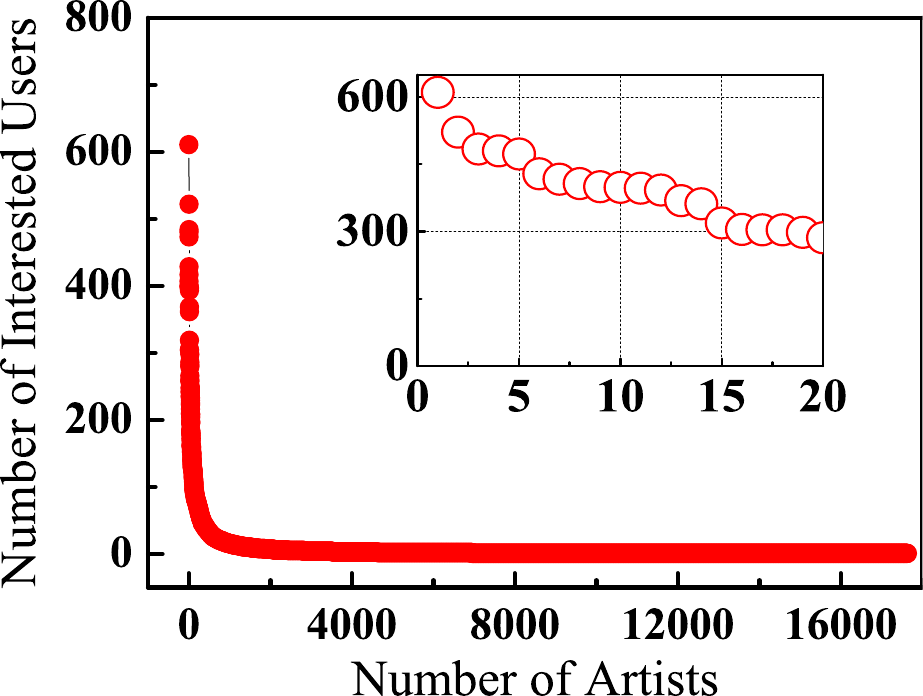}}
\subfigure[\small Tag distribution in terms of interested user amount]{
\includegraphics[height=1.2in]{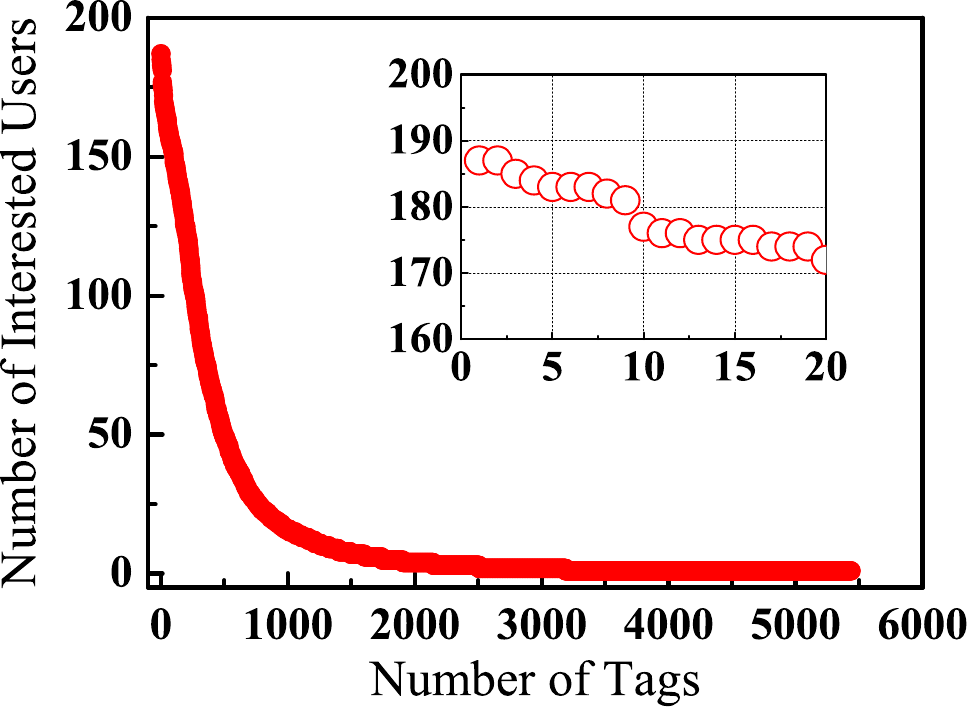}}
\caption{Statistics of the content retrieving trace}
\label{fig:dataset}
\end{figure}

Then, in order to simulate named data in content servers, we select a content retrieving trace\footnote{Last.fm dataset. http://grouplens.org/datasets/hetrec-2011/.}, \emph{Last.fm} which tracks 1892 users listening habits of music in terms of 17632 artists and 5429 music-tags.
The statistic as shown in Fig. \ref{fig:dataset} reflects that both artist and tag popularity follow the power-law distribution, which is in accordance with the content popularity distribution \cite{cha2009analyzing} mentioned in section \ref{sec:introduction}. Therefore,
to facilitate the simulation, we consider a named data is uniquely indicated by a name ``/artist/music-tag", and we can extract the popular
artists and music-tags in the trace to build the global namespace $N$. Specifically, we select top 10 popular artists (i.e., over 20\% users are interested in them) and randomly select 10 common music-tags in terms of those 10 artists (there are 15 common music-tags in total).
In other words, there are 100 distinct named popular data in content servers. We should emphasize that those numbers are a moderate setting. For example, higher values may result in weak social strength and thus it is difficult for users to form their Friendship Circles.
For each user, the \emph{Last.fm} trace also reflects the interest preference of different artists and music-tags according to the listening record history. For example, if 34\% of songs a user listened are ``pop", we consider that she has 34\% preference to that music-tag. In accordance to the size of users in user contact traces, we randomly select several user profiles related to those popular artists and music-tags as user interest preference.

Finally, we randomly select 5 Access Points in \emph{Infocom06} (5 Cellular Towers in \emph{MIT Reality}) to simulate base stations.
We consider that each mobile user generates an Interest packet by her own preference every 10 minutes. If the user is connecting to a base-station she will obtain the desired data directly with probability 50\% (The rest 50\% is to model the users who are sensitive to the monetary
cost of cellular data usage), and otherwise she launches the cooperative content retrieval. The first day of user contact traces is used for experimental warm-up such as accumulating contacts and contents storage without cooperative content retrieval (In the warm-up stage, the content request TTL is fixed to 2 hours, and a user can obtain requested contents directly with probability 1 if she connects a base station within TTLs).
\begin{figure*}[thb]
\centering
\subfigure{
\includegraphics[height=1.4in]{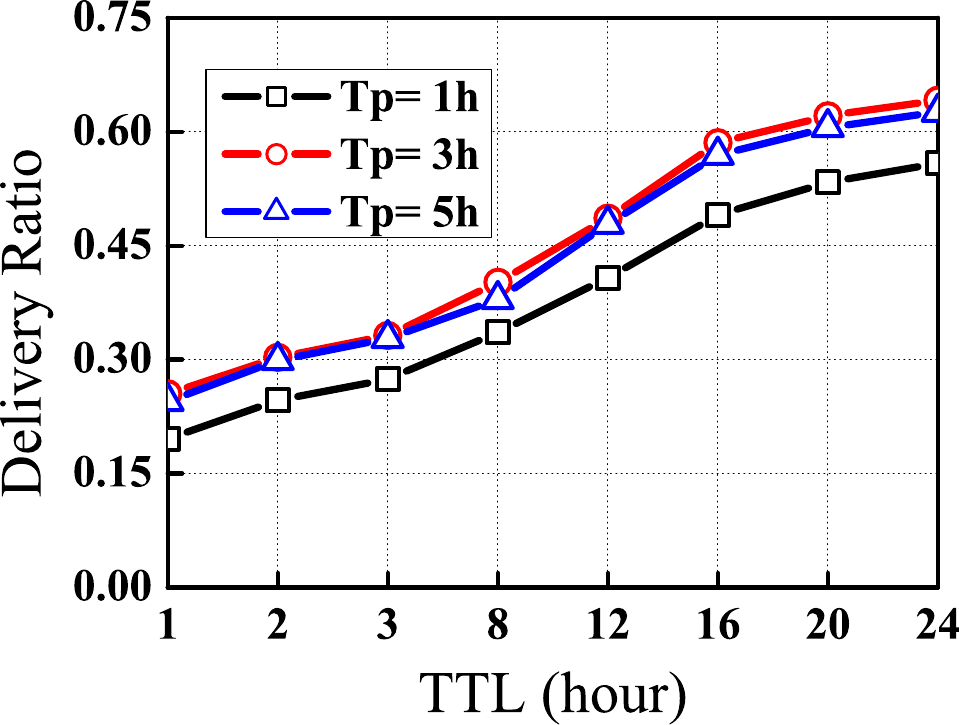}}
\subfigure{
\includegraphics[height=1.4in]{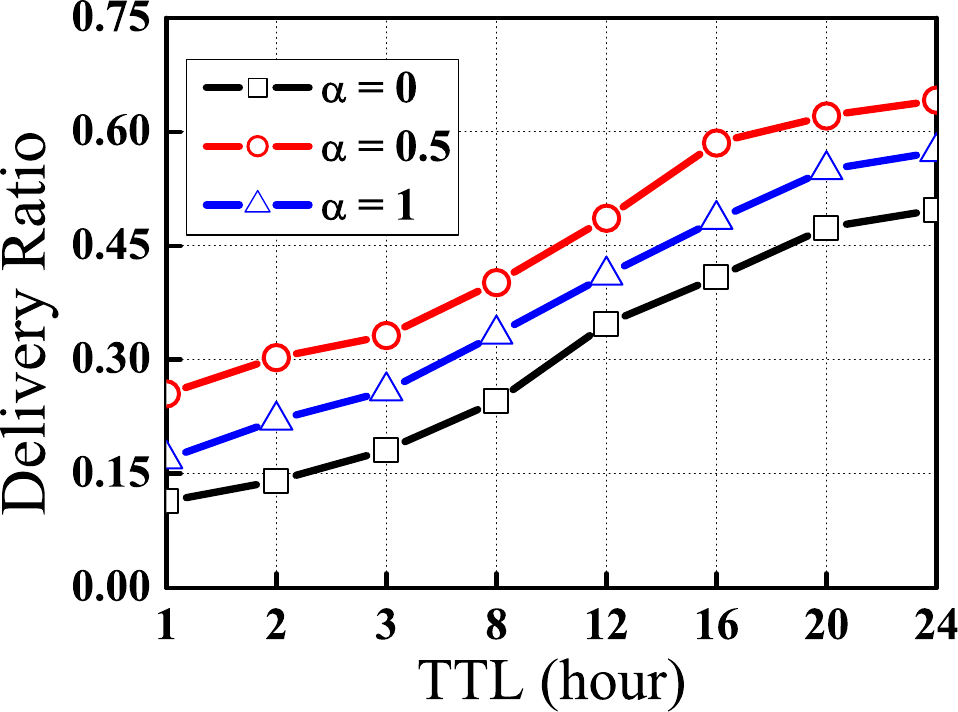}}
\subfigure{
\includegraphics[height=1.4in]{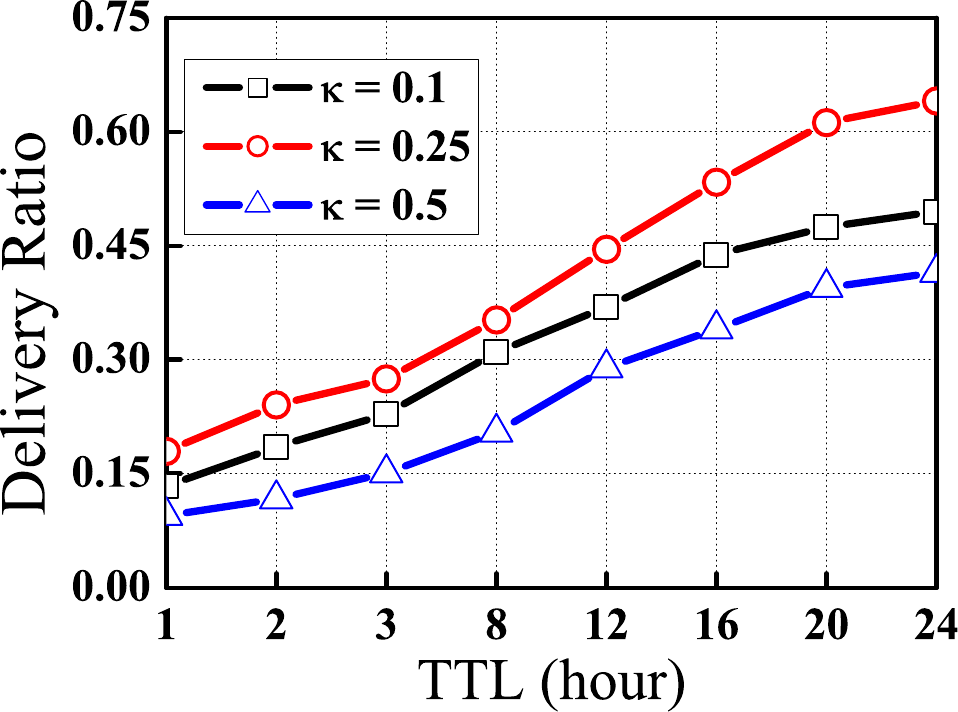}}
\caption{The impact of different parameters ($T_p$, $\alpha$ and $\kappa$) on \emph{Infocom06} trace}
\label{fig:info_para}
\end{figure*}
\begin{figure*}[thb]
\centering
\subfigure{
\includegraphics[height=1.4in]{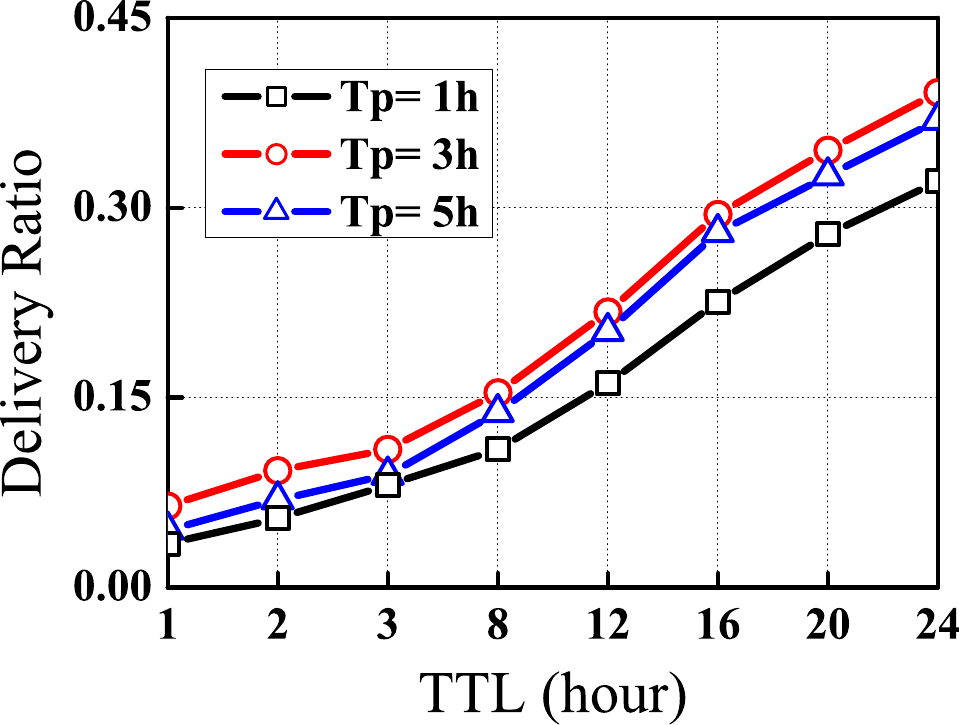}}
\subfigure{
\includegraphics[height=1.4in]{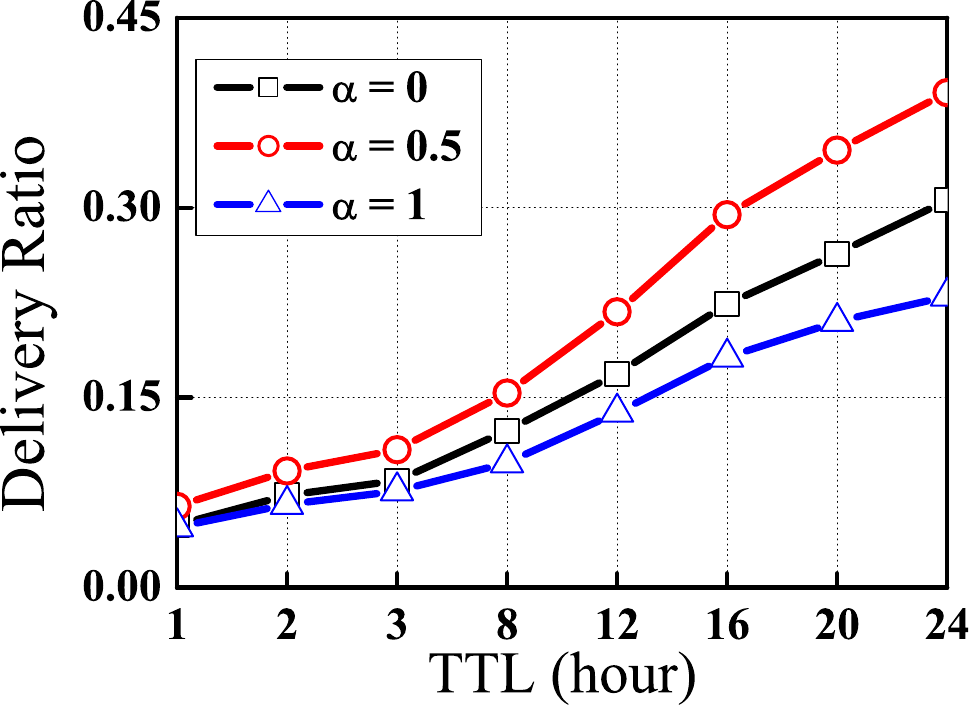}}
\subfigure{
\includegraphics[height=1.4in]{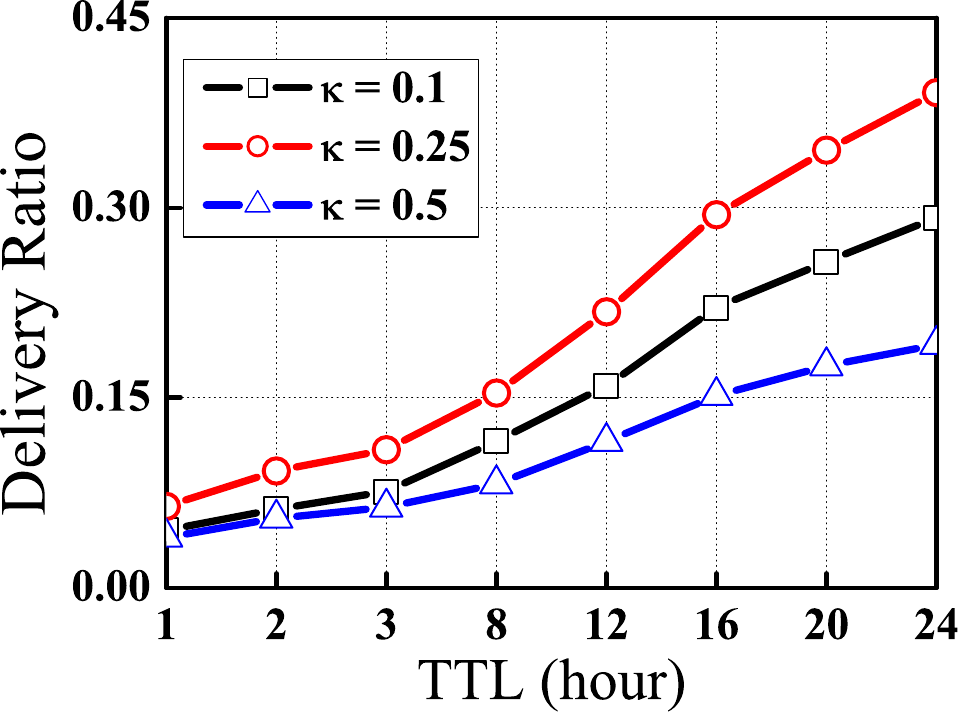}}
\caption{The impact of different parameters ($T_p$, $\alpha$ and $\kappa$) on \emph{MIT Reality} trace}
\label{fig:mit_para}
\end{figure*}

\emph{2) Simulation Setup}:
We implement the scenario and the sNDN framework in the Opportunistic Networking Environment (ONE) simulator\footnote{http://www.netlab.tkk.fi/tutkimus/dtn/theone/.}.
For our evaluation, we measure the impact of the time window $T_p$, strength ratio $\alpha$ and key name component threshold $\kappa$ on the sNDN performance, and then compare sNDN with
\begin{itemize}
  \item Flood, a flooding data searching and delivery method which is viewed as the upper bound in terms of delivery ratio and delay;
  \item STCR \cite{lu2014information}, an advertisement-based content retrieval method as mentioned in section \ref{sec:related};
  \item FC-BubbleRap, a revised BubbleRap method \cite{hui2011bubble} for cooperative content retrieval, which takes Named Friendship Circle to substitute community structure;
  \item Direct, a naive method which the user herself searches for the desired data without any replication.
\end{itemize}

The performance is measured with three metrics: \emph{delivery ratio}, the ratio between the number of delivered packets (i.e, Data packets) to
the total number of created ones (i.e., Interest packets), which implicitly indicates the capacity-saving degree of cellular network. For example, 80\% delivery ratio means that 80\% data requests can be fulfilled by nearby users via D2D communications; \emph{actual delay}, average delay of all delivered packages and \emph{overhead}, average
number of relays used for one delivered packet. We run the simulation 50 times where the user interest profile is randomly selected in each run, and adopt the average results to depict the following graphs. In addition, we should emphasize that, due to the limited user amount and user contacts in user contact traces, the TTL setting in the simulation is not to reflect the practical requirements but to test what the performance upper-bound of a routing scheme will be and how the performance gap among different routing schemes, which shares the same spirit with many existing works\cite{ hui2011bubble, gao2011user, lin2012preference}.

\emph{3) Simulation Results}:
We conduct experiments to evaluate the parameter $T_p$, $\alpha$ and $\kappa$. Note that, our purpose is not to find the optimal values of the parameters but to shed light on the effect of them on the performance. Therefore, we only compare representative values and the results are shown in Fig. \ref{fig:info_para} and \ref{fig:mit_para}. The y-axis in the figures represents the delivery ratio metric and the x-axis represents the request TTL before which the returned Data packets are valid (i.e., the countdown timer mentioned in section \ref{sec:scenario}).
We first discuss the impact of windows size $T_p$ which determines how long the history should be retrospected. The result when $T_p$ is 1h is the worst one in two figures. This is because the value of social strength is too small regarding limited history records, which is adverse to the construction of Named Friendship Circle. With $T_p$ increasing, the performance is improving. Besides, the remaining two results share close performance when $T_p$ increases. This situation indicates that the formulation of social strength emphasizes the importance of recent records and it is no need to consider many stale records. That is, a small value of $T_p$ (e.g., 3h) is sufficient.
Then, we discuss the impact of ratio $\alpha$ which determines the different weights on physical strength and logical strength.
the results show that the combination of both physical and logical strength achieves much better performance. Specifically, in Fig. \ref{fig:info_para}, the result when $\alpha$ is 0.5 achieves over 20\% improvement compared to the other two results and the peak value is 50\% and 70\% better than that using single strength in Fig. \ref{fig:mit_para}. The reason is that although physical strength can build Friendship Circle with the mobility-centric friends, these friends may not share common data, which will fuzz up the ``data area''. Besides, although the logical strength can well indicate the ``data area'', the performance will suffer from low contact frequency among friends. Therefore, the social strength including both of them is more appropriate. Besides, we observe that if the network is dense, it is better to consider logical strength more (i.e., the performance when $\alpha$ is 1 is better than that is 0).
Finally, we discuss the impact of threshold $\kappa$ which determines the name size of neighbor set (also the name size of Friendship Circle). The result when $\kappa$ is 0.5 is the worst one in two figures. The reason is that a higher threshold value will shrink the user Friendship Circle name, and hinders the first step of Interest packet routing (i.e., quickly find the desired Friendship Circle). However, a lower threshold value enables a Friendship Circle name to incorporate more name components that many users in the circle may not have the data associated to them, and hence hinders the second step of Interest packet routing (i.e., quickly find the desired data). As such, we can see that a moderate value (i.e., $\kappa$=0.25) is preference, which achieves more than 30\% performance increase than the others in two figures.

\begin{figure*}[!thb]
\centering
\subfigure[\small Delivery ratio]{
\includegraphics[height=1.4in]{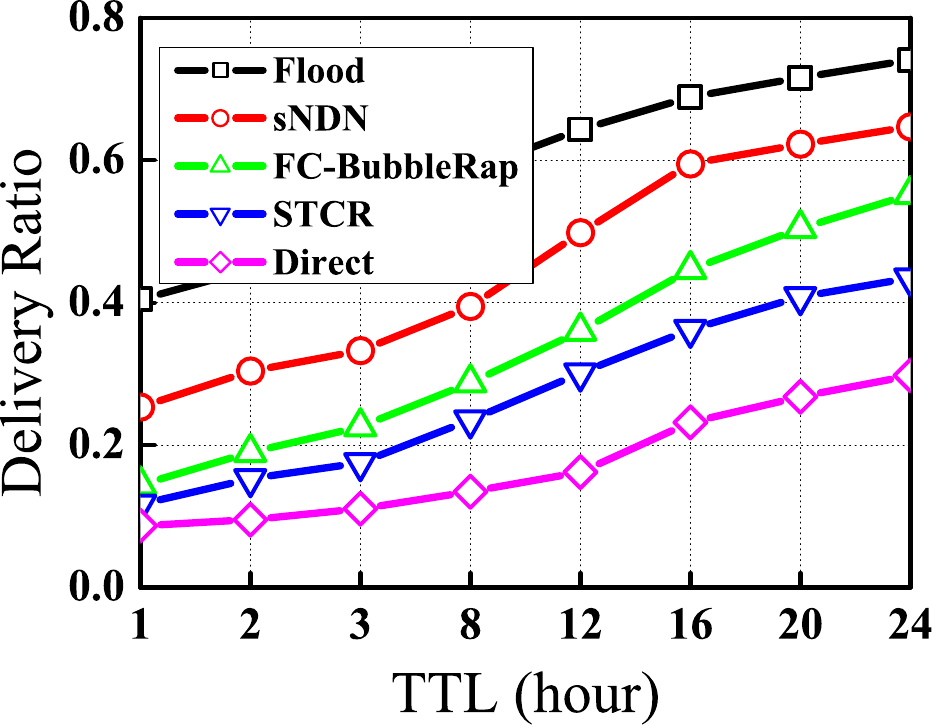}}
\subfigure[\small Actual delay]{
\includegraphics[height=1.4in]{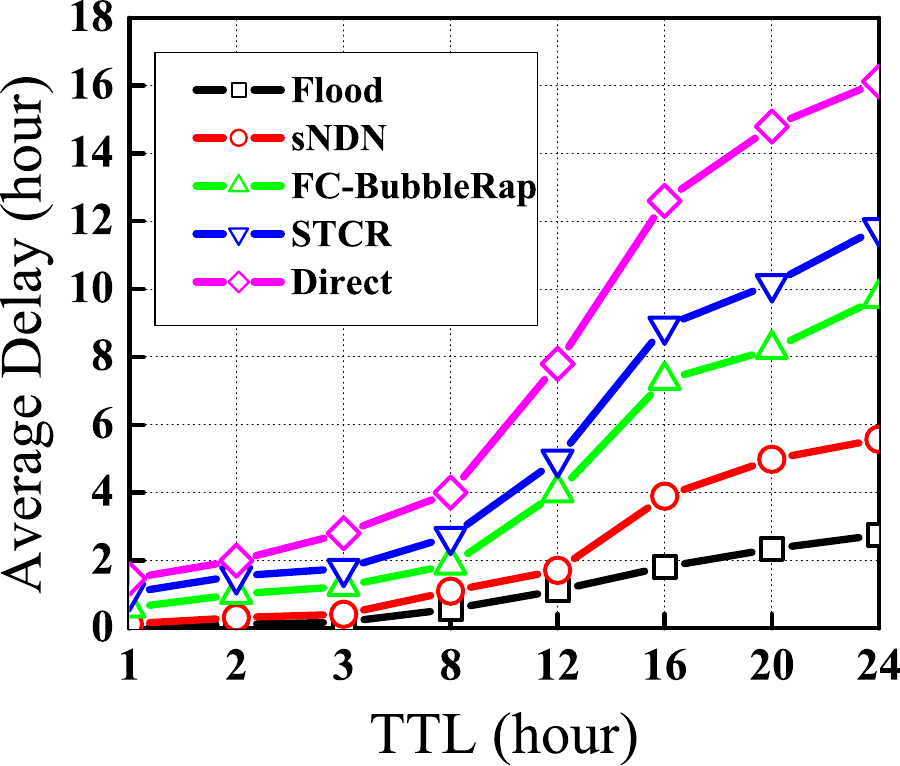}}
\subfigure[\small Overhead]{
\includegraphics[height=1.4in]{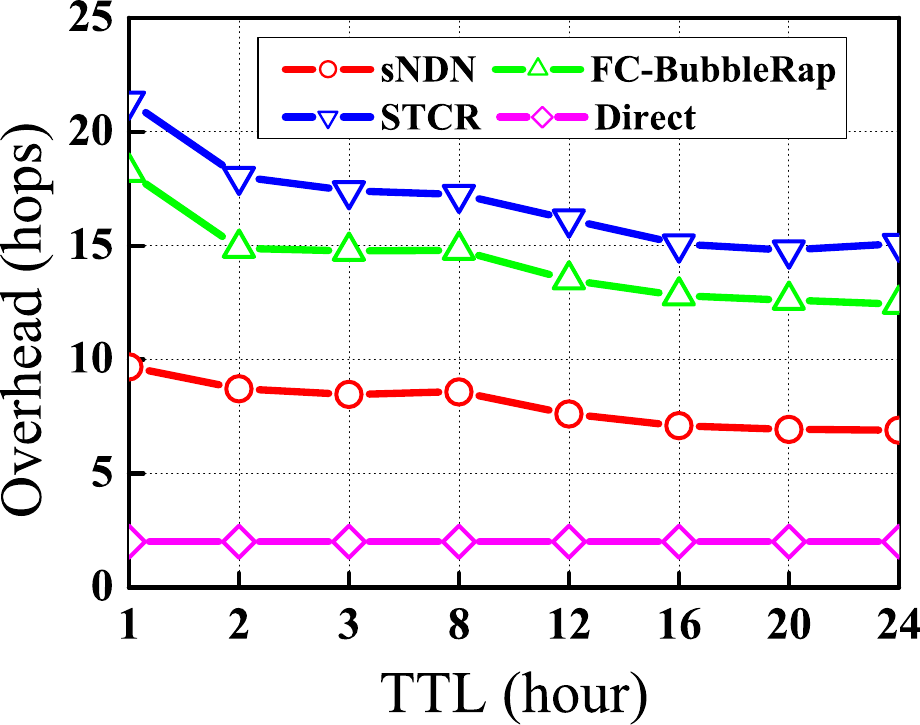}}
\caption{The overall content retrieval performance of different schemes on \emph{Infocom06} trace}
\label{fig:info}
\end{figure*}

\begin{figure*}[!thb]
\centering
\subfigure[\small Delivery ratio]{
\includegraphics[height=1.4in]{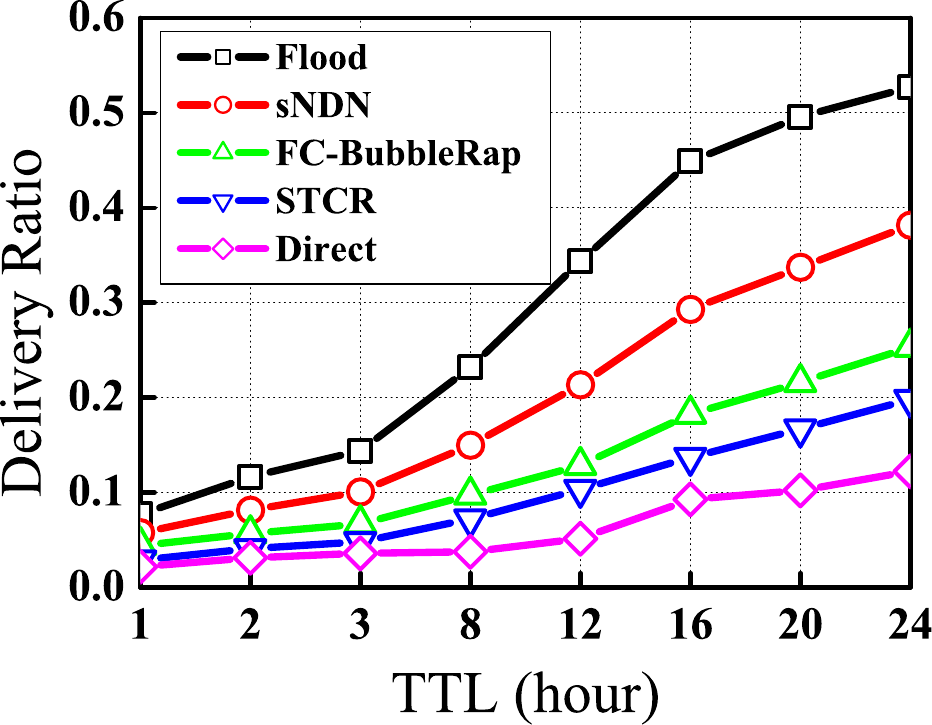}}
\subfigure[\small Actual delay]{
\includegraphics[height=1.4in]{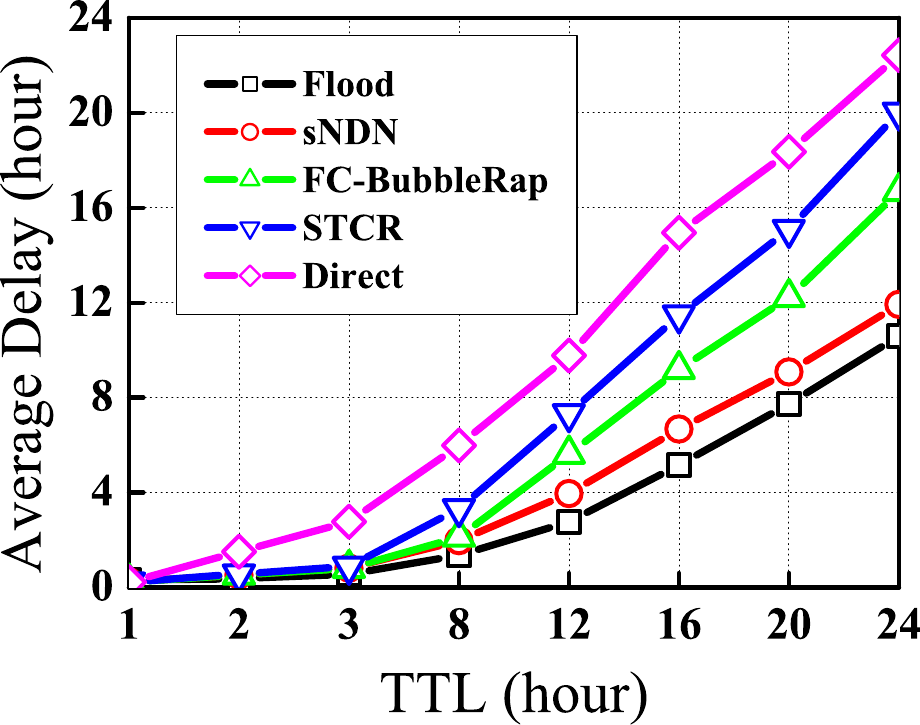}}
\subfigure[\small Overhead]{
\includegraphics[height=1.4in]{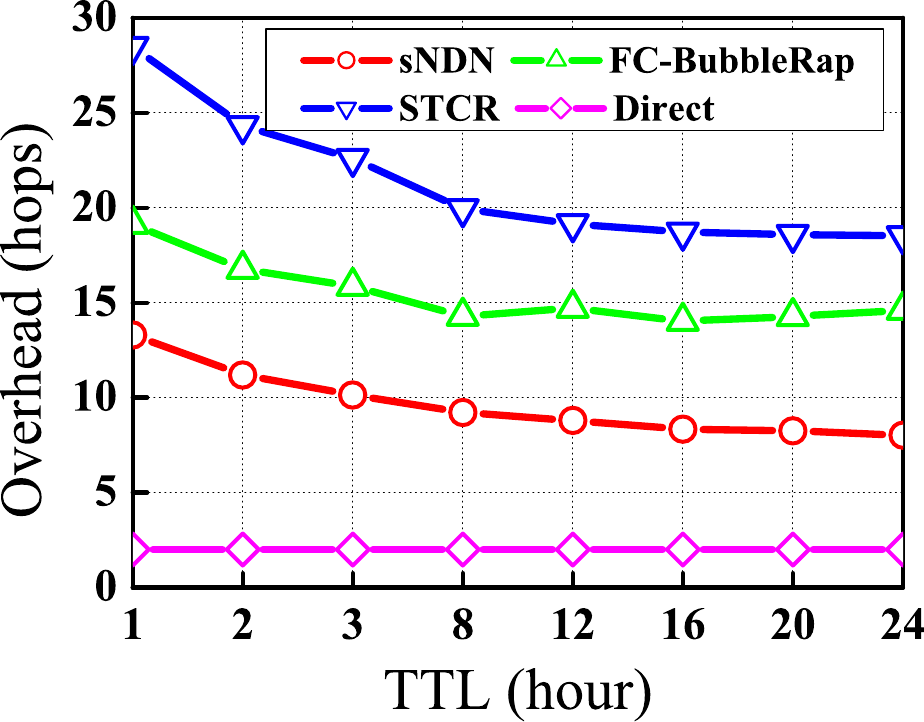}}
\caption{The overall content retrieval performance of different schemes on \emph{MIT Reality} trace}
\label{fig:mit}
\end{figure*}

Fig. \ref{fig:info} and Fig. \ref{fig:mit} depict the performance of different content retrieval schemes. The y-axis in the figures represents the preceding three metrics and the x-axis represents the request TTL before which the returned Data packets are valid. Generally speaking, the performance of sNDN is better than all the non-flooding protocols and achieves comparable results to the upper bound. Specifically, in the dense environment (Fig. \ref{fig:info}), since cooperative users are sufficient, the delivery ratio of sNDN is only 14.8\% less than that of Flood when the TTL is 24 hours, and achieves 20.5\% improvement compared to FC-BubbleRap on average. Besides, with the help of Named Friendship Circle, not only sNDN but also FC-BubbleRap are 48.9\% and 27.1\% better than STCR. The actual delay shares the same trend, and sNDN only requires less than half of time compared with all the non-flooding protocols.
In the spare environment (Fig. \ref{fig:mit}), with more cooperative users, the delivery ratio of Flood can achieve 38\% gain above ours when the TTL is 24 hours. Even though, the delivery ratio of sNDN still achieves 50.3\% improvement over FC-BubbleRap and doubles the performance of the other two schemes.
As to the actual delay, sNDN outperforms other non-flooding protocols over 25\% in the end. Fig. \ref{fig:info}(c) and Fig. \ref{fig:mit}(c) present the overhead performance, and the sNDN keeping 8$\sim$12 relays has much less cost. In addition, the evaluation results demonstrate that adopting sNDN for cooperative content retrieval can save cellular capacity greatly (i.e., roughly 60\% for dense network and 40\% for sparse network).


\section{Conclusion}
In this paper, we advocate \emph{Content Retrieval At the Edge}, a content-centric cooperative service paradigm via device-to-device (D2D) communications to reduce cellular traffic volume in mobile networks. As the first step to such a cooperative content retrieval, we design \emph{sNDN}, a social-aware named data framework. Specifically, sNDN first introduces Friendship Circle by grouping a user with her close friends of high mobility similarity and content similarity. Then, it constructs NDN routing tables conditioned on Friendship Circle encounter frequency to navigate a content request and content reply packet between Friendship Circles, and leverage social properties in Friendship Circle to search for the final target as inner-Friendship Circle routing. The evaluation results also demonstrate that adopting sNDN for cooperative content retrieval can save cellular capacity greatly and sNDN outperforms other content retrieval schemes significantly. In the future, besides taking optimal forwarder selection into account we will extend sNDN with a operator-assisted feature, and further explore the content cache issue to facilitate the Content Retrieval At the Edge.

%\section{Acknowledgment}
%This work is supported in part by the EU FP7 IRSES MobileCloud project, Simulation Center (Lower Saxony Ministry for Culture and Education), and the Natural Science Foundation of Tianjin, China (No.16JCQNJC00700).
%
\bibliographystyle{IEEEtran}
\bibliography{TETC_citefile}

\end{document}